\documentclass[12pt, journal,onecolumn]{IEEEtran} 

\usepackage{amssymb,amsmath,epsfig}

\usepackage{cite}
\usepackage{psfig}
\usepackage{algorithmic}
\usepackage{algorithm}
\usepackage{amsfonts}
\usepackage{amstext}
\usepackage{graphicx}
\usepackage{setspace}
\usepackage{amssymb}
\usepackage{esint}
\usepackage{amsthm}
\usepackage{perpage} 
\usepackage{epstopdf}

\renewcommand{\baselinestretch}{2}

\begin{document}

\title{Identification of SM-OFDM and AL-OFDM Signals Based on Their Second-Order Cyclostationarity}


\author{\IEEEauthorblockN{Ebrahim~Karami,~\IEEEmembership{Member,~IEEE}
and Octavia~Dobre,~\IEEEmembership{Senior~Member,~IEEE}
\\\IEEEauthorblockA{Department of Engineering and Applied Sciences, Memorial University of Newfoundland, St. John's, Canada}
\\\IEEEauthorblockA{Email: \{ekarami,odobre\}@mun.ca}}}


\maketitle
\IEEEpeerreviewmaketitle

\begin{abstract}
Automatic signal identification (ASI) has important applications to both commercial and military communications, such as software defined radio, cognitive radio, spectrum surveillance and monitoring, and electronic warfare. While ASI has been intensively studied for single-input single-output systems, only a few investigations have been recently presented for multiple-input multiple-output systems. This paper introduces a novel algorithm for the identification of spatial multiplexing (SM) and Alamouti coded (AL) orthogonal frequency division multiplexing (OFDM) signals, which relies on the second-order signal cyclostationarity. Analytical expressions for the second-order cyclic statistics of SM-OFDM and AL-OFDM signals are derived and further exploited for the algorithm development. The proposed algorithm provides a good identification performance with low sensitivity to impairments in the received signal, such as phase noise, timing offset, and channel conditions.
\end{abstract}

\begin{keywords}
Automatic signal identification (ASI), multiple-input multiple-output (MIMO), space-time block code (STBC), orthogonal frequency division multiplexing (OFDM), cyclostationarity, cyclic correlation function (CCF), cycle frequency (CF).
\end{keywords}

\section{Introduction}

Automatic signal identification (ASI) has various applications to both commercial and military communications, such as software defined radio, cognitive radio, spectrum surveillance and monitoring, and electronic warfare \cite{classf_m2,gardner1992signal,Dobre1,software_radio2,Cabric,Cognitive_Radio_survey2009}. 
While ASI was originally applied to military communications \cite{classf_m2,gardner1992signal}, recent developments and trends in commercial wireless communications have allowed its introduction in the context of commercial software defined and cognitive radios \cite{Dobre1,software_radio2,Cabric,Cognitive_Radio_survey2009}. 
With the software defined radio, the parameters of the transmitted signal, e.g., antenna configuration and modulation format, are adjusted according to the environment; hence, their blind identification/estimation is required at the receiver. Furthermore, a main task of cognitive radios is spectrum awareness, which enables the detection and identification of the existing signals to facilitate transmission with acceptable interference. ASI represents a challenging problem under the conditions of no a priori knowledge of the transmitted data and signal parameters, as well as channel effects. The aim is to devise ASI algorithms which do not rely on pre-processing, e.g., do not require channel estimation, and provide an acceptable identification performance at lower signal-to-noise ratios (SNRs) and within a short observation interval. 
 The majority of the ASI literature is devoted to single-input single-output systems, focusing on the identification of various modulation schemes \cite{Dobre1,Wu_MOD_Classification_2008,WeiSu,classf_m2,Grimaldi,Bjorsell,Ozdemir,Dobre2,cl_Sc5}, single-carrier versus multi-carrier transmissions \cite{oner2007extraction,Dobre_MOD_Cyclostationarity,Zhang_TW}, and different multi-carrier transmissions \cite{ciblat1,cyclo1,Alaa3}.
Recently, the ASI problem has started to be investigated for multiple-input multiple-output (MIMO) systems, mainly due to their adaptation by various wireless communications standards, such as IEEE 802.16e, 3GPP LTE, and IEEE 802.11n \cite{book_lte,wimax_1,WIMAX}. 
There is only a relatively small number of papers tackling the ASI problem for MIMO systems, as follows: the estimation of the number of transmit antennas was studied in \cite{No_Trans_anrennas1,No_Trans_anrennas2}, modulation classification was investigated in \cite{MOD_MIMO_Choqueuse,Muhl1,Muhl2,Haimovich}, and identification of space-time block codes (STBCs) was explored in  \cite{choquese1,choquese2,choquese55,Deyoung,Miao56,Marey_cyclost,Yahia_TCOM,Mar_Dob}.
Regarding the STBC identification, the maximum likelihood approach was studied in \cite{choquese1}, while the feature-based approach was considered in \cite{choquese2,choquese55,Deyoung,Miao56,Marey_cyclost,Yahia_TCOM,Mar_Dob}. Although the former provides an optimal solution in the sense of maximizing the average probability of correct identification, it suffers from an exponential computational complexity and requires knowledge of the
channel parameters, as well as symbol and block synchronization \cite{choquese1}. With the latter approach, features are extracted from the received signal and a decision is made based on their difference. 
In \cite{choquese2,choquese55}, the Frobenius norm of the time-lag correlation function was exploited as a discriminating feature for different STBCs, and identification was performed based on a binary tree algorithm \cite{choquese2} and on minimizing the distance between
the theoretical and estimated Frobenius norm \cite{choquese55}. Fourth- and
second-order cyclostationarity was respectively employed in \cite{Deyoung} and \cite{Miao56} to distinguish between spatial multiplexing (SM) and Alamouti (AL) codes. Furthermore, second-order cyclostationarity was investigated in \cite{Marey_cyclost} to identify different STBCs in the presence of transmission impairments. The fourth-order moment and the discrete Fourier transform of the fourth-order lag product were employed in \cite{Yahia_TCOM} to identify SM and AL codes. 
The previously reported researches have considered single-carrier transmissions over frequency-flat channels. However,
in practice, high data rate applications involve transmission
over broadband frequency-selective channels, for which MIMO-orthogonal frequency division multiplexing (OFDM) technology provides an efficient solution \cite{OFDM_survey}. To the best of our knowledge, there exist two papers in the literature that address the ASI problem for MIMO-OFDM systems \cite{Haimovich,Mar_Dob}. Modulation classification for SM signals is investigated in \cite{Haimovich}, while identification of STBCs is studied in \cite{Mar_Dob}, with the latter being relevant to our work. We propose a novel identification algorithm for the identification of SM-OFDM and AL-OFDM signals; this relies on the second-order cyclostationarity of the received signal, whereas the algorithm in \cite{Mar_Dob} employs signal moments. We derive the analytical expressions for the cyclic cross-correlation function (CCF) and its corresponding cycle frequencies (CFs) for the SM-OFDM and AL-OFDM signals, and then use them to develop a novel feature-based signal identification algorithm.
The rest of the paper is organized as follows. Section \ref{sec:System-Model}
presents the system model and Section \ref{sec:Cycl} introduces the analytical results for the CCF and CF, as well as the identification algorithm. Simulation results are showed in Section \ref{sec:sims} and conclusions are drawn in Section \ref{sec:Conclusion}.

\section{System Model}\label{sec:System-Model}
\begin{figure}
\centering
\includegraphics[width=1\linewidth]{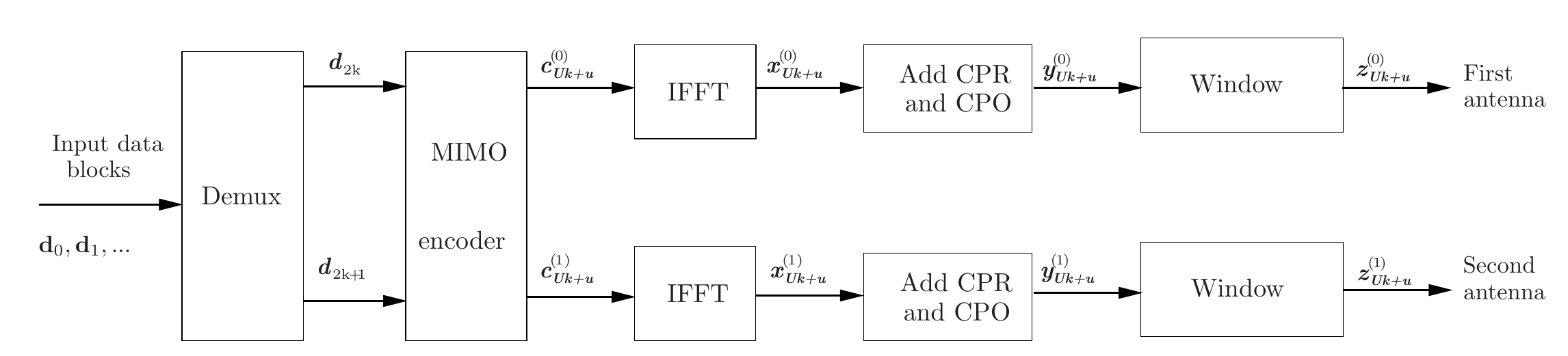}
\caption{Block diagram of a MIMO-OFDM transmitter.}
\label{Block_MIMO}
\end{figure}

The baseband equivalent block diagram of a MIMO-OFDM transmitter is presented in Fig. \ref{Block_MIMO}. 
The input signal is a stream of data blocks, $\mathbf{d}_{t}=\left[d_t(0), d_t(1), \cdots, d_t(N-1)\right]$, where each block contains $N$ independent and identically distributed (i.i.d.) symbols drawn from either an $\Omega$-ary quadrature amplitude modulation (QAM) or phased-shift-keying (PSK) signal constellation, $\Omega \ge 4$. 
Two transmit antenna ($N_t=2$) are considered, and, accordingly, the data stream is demultiplexed into two sub-streams. Such sub-streams are fed into the MIMO encoding block, which in this work is either SM or AL. Hence, the $k$th group of two data blocks, $\left(\mathbf{d}_{2k},\mathbf{d}_{2k+1}\right)$, is encoded according to a code matrix $\mathbf{C}\left(\mathbf{d}_{2k},\mathbf{d}_{2k+1}\right)$ of size $N_t \times UN$, in order to be transmitted during $U$ block instants \cite{AL_two}. Note that $U=1$ for SM and $U=2$ for AL.   
The code matrices $\mathbf{C}^{(\textrm{SM})}$ and $\mathbf{C}^{(\textrm{AL})}$ corresponding to the SM\footnote{Note that $N_t=2$ is considered for SM; however, the identification algorithm is applicable for $N_t > 2$, as well.} and AL encoders are respectively given by \cite{Mar_Dob,AL_two}

\begin{equation}
\mathbf{C}^{(\textrm{SM})}\left(\mathbf{d}_{2k},\mathbf{d}_{2k+1}\right)=\left[\begin{array}{cc}
\mathbf{c}_{1k+0}^{(0)}\\
\mathbf{c}_{1k+0}^{(1)}
\end{array}\right]=\left[\begin{array}{cc}
\mathbf{d}_{2k}\\
\mathbf{d}_{2k+1} \end{array}\right], \label{eq:SM}\end{equation}
\noindent and
\begin{equation} 
\mathbf{C}^{(\textrm{AL})}\left(\mathbf{d}_{2k},\mathbf{d}_{2k+1}\right)=\left[\begin{array}{cc}
\mathbf{c}_{2k+0}^{(0)} & \mathbf{c}_{2k+1}^{(0)}\\
\mathbf{c}_{2k+0}^{(1)} & \mathbf{c}_{2k+1}^{(1)}
\end{array}\right]=\left[\begin{array}{cc}
\mathbf{d}_{2k} & -\mathbf{d}_{2k+1}^*\\
\mathbf{d}_{2k+1}& \mathbf{d}_{2k}^* \end{array}\right], \label{eq:AL}\end{equation}

\noindent where $\mathbf{c}_{Uk+u}^{(f)}$ represents the data block to be transmitted from the $f$th antenna, $f=0,1$, at block instant $Uk+u$, $u=0,...,U-1$, and * denotes complex conjugation. 
 
 The output of the MIMO encoder is fed into the inverse fast Fourier transform (IFFT) block, yielding the OFDM symbol $\mathbf{x}_{Uk+u}^{(f)}=[x_{Uk+u}^{(f)}(0),x_{Uk+u}^{(f)}(1),...,x_{Uk+u}^{(f)}(N-1)]$ as
   
 \begin{equation}\label{IFFT}
 x_{Uk+u}^{(f)}(n)=\frac{1}{\sqrt{N}}\sum\limits_{n_1=0}^{N-1}c_{Uk+u}^{(f)}(n_1)e^{j\frac{2 \pi n n_1}{N}},\quad n=0,1,...,N-1,
\end{equation}
 
\noindent where $c_{Uk+u}^{(f)}(n_1)$ is the $n_1$th element of $\mathbf{c}_{Uk+u}^{(f)}$ , $n_1=0,1,...,N-1$.

 The cyclic prefix (CPR), which represents a copy of the last $N_G$ samples of the OFDM symbol, is then added. Windowing is also applied; this increases the CPR to $\nu=N_G+N_W$, where $N_W$ is the number of samples in the transition time between two consecutive OFDM symbols \cite{Prasad}. Furthermore, the first $N_W$ samples of the OFDM symbol are transmitted after the effective part of the symbol, during the next transition time, as a cyclic postfix (CPO) \cite{Prasad}. By taking into account the CPR, CPO, and windowing, the OFDM symbol is expressed as $\mathbf{z}_{Uk+u}^{(f)}=[z_{Uk+u}^{(f)}(-\nu),z_{Uk+u}^{(f)}(-\nu+1),...,z_{Uk+u}^{(f)}(N+N_W-1)]$, with
  
\begin{equation}\label{eq:Win}
z_{Uk+u}^{(f)}(n)=W_n x_{Uk+u}^{(f)}(\tilde{n}), \quad n=-\nu,...,N+N_W-1, \quad f=0,1, \quad u=0,...,U-1,
\end{equation}
where $W_n$, with $n=-\nu,...,N+N_W-1$, represent the window coefficients\footnote{Note that the commonly used raised-cosine window is considered in this work.} \cite{Prasad} and $\tilde{n}=mod(n,N)$.

Finally, the transmit sequence $\{s^{(f)}(m)\}$ from the $f$th antenna, $f=0,1$, is expressed as
\begin{equation}\label{Tr_sig}
s^{(f)}(m)=\sum\limits_{k=-\infty}^{+\infty} \sum\limits_{u=0}^{U-1}\sum\limits_{n=-\nu}^{N+N_W-1}z_{Uk+u}^{(f)}(n)\delta(m-(Uk+u)(N+\nu)-n), 
\end{equation}
\noindent where $\delta(m)$ is the Kronecker
delta function equal to one if $m=0$ and zero otherwise.
  The transmit sequence $\{s^{(f)}(m)\}$ from the $f$th antenna propagates through an unknown frequency-selective wireless channel.
Hence, the $m$th sample of the signal
received at the $v$th receive antenna, $r^{(v)}(m)$, can be expressed as 
\begin{equation}
r^{(v)}(m)=\sum\limits_{f=0}^{1}\sum\limits_{p=0}^{L_{p}-1}h_{vf}(p)s^{(f)}(m-\vartheta(p))+w^{(v)}(m),\label{eq:received_sample}\end{equation}
\noindent where $L_p$ is the number of propagation paths, $h_{vf}(p)$ is the channel coefficient corresponding to the $p$th path between the $f$th transmit and $v$th receive antenna, $\vartheta(p)$ is the propagation delay corresponding to the $p$th path, and $w^{(v)}(m)$ is the additive white Gaussian noise (AWGN) with variance $\sigma_w^2$. Subsequently, we will develop an algorithm to identify the SM-OFDM and AL-OFDM signals from the received sequences $\left\{ r^{(v)}(m)\right\}$, $v=0,1,...,N_r-1$, where $N_r$ is the number of receive antennas.
\section{Second-Order Cyclostationarity-Based SM-OFDM and AL-OFDM Signal Identification \label{sec:Cycl}}
In this section, the CCF and its corresponding CFs are derived for the SM-OFDM and AL-OFDM signals and employed to develop a novel feature-based signal identification algorithm.

\subsection{Cyclostationarity Concept}
The received sequences $\{r^{(0)}(n)\}$ and $\{r^{(1)}(n)\}$\footnote{Note that the case of two receive antennas ($N_r=2$) is considered; later in the paper, $N_r > 2$ will be also studied.} exhibit second-order cyclostationarity if their first\footnote{Note that due to the symmetry in the signal constellations, the first-order statistics equal zero.} and second-order time-varying correlation functions are periodic in time \cite{gardner1994cumulant}. Here we consider the non-conjugate second-order time-varying cross-correlation function, defined as
\begin{equation}\label{eq:corr}
 c(m,\tau)=\mathrm{E}\left[r^{(0)}(m)r^{(1)}(m+\tau)\right],
\end{equation}
\noindent where $\mathrm{E}\left[.\right]$ is the statistical expectation and $\tau$ is the delay. If $c(m,\tau)$ is periodic in $m$ with the fundamental period $M_0$, then it can be expressed by a Fourier series \cite{gardner1994cumulant},
\begin{equation}
c(m,\tau)=\sum\limits_{\{\alpha\}}C(\alpha,\tau)e^{j2\pi m \alpha},\label{eq:Fourier_series}\end{equation}
\noindent where the coefficients
 \begin{equation}
C(\alpha,\tau)=\frac{1}{M_0}\sum\limits_{m}c(m,\tau)e^{-j2\pi m \alpha},\label{eq:CCF}
 \end{equation}
\noindent are referred to as the CCF at CF $\alpha$ and delay $\tau$, and the set of CFs is given by $\{\alpha\}=\{\ell/M_0, \, \ell \in \mathbb{\mathcal{\mathcal{I}}} \, \mbox{, with} \, \mathbb{\mathcal{\mathcal{I}}} \, \mbox{as the set of integers}\}$.
\subsection{Analytical Expressions for the CCF of the SM-OFDM and AL-OFDM Signals}
The analytical expressions for the CCF and its corresponding CFs are derived here for the SM-OFDM and AL-OFDM signals. Results are obtained by following the commonly used assumptions that \cite{choquese1,choquese2,choquese55,Deyoung,Miao56,Marey_cyclost,Yahia_TCOM,Mar_Dob}: a) the transmitted sequences are uncorrelated with the noise: $\mathrm{E}\left[s^{(f)}(m_0)w^{(v)}(m_1)\right]=0$, $\forall\, $ $f,v=0,1$, and $m_0,m_1\in\mathbb{\mathcal{\mathcal{I}}}$; b) the noise in each channel is uncorrelated with that of the
other channels: $\mathrm{E}\left[w^{(v_0)}(m_0)w^{(v_1)}(m_1)\right]=$$\mathrm{E}\left[w^{(v_0)}(m_0)(w^{(v_1)}(m_1))^*\right]=0$,
$\forall\,$ $m_0,m_1\in\mathbb{\mathcal{\mathcal{I}}}$,  $v_0,v_1=0,1$, 
and $v_0\neq v_1$; c) the data symbols are uncorrelated with each other: $\mathrm{E}\left[d_{k_0}(n_0)d_{k_1}(n_1)\right]=0$, $\mathrm{E}\left[d_{k_0}(n_0)d_{k_1}^*(n_1)\right]=$$\sigma_{s}^{2}\delta(k_0-k_1)\delta(n_0-n_1)$,
$\forall$ $k_0$, $k_1$, $n_0$ and $n_1$, where
$\sigma_{s}^{2}$ is the transmit signal power; and d) the channel gains for each transmit-receive
antenna link remain constant over the observation interval.
\begin{flushleft}
\textbf{SM-OFDM}
\par\end{flushleft}
By using (\ref{eq:SM}), (\ref{IFFT}), and (\ref{eq:Win}), one can easily show that
\par
\begin{equation}
\mathrm{E}\left[z_{k_0}^{(f_0)}\left(n_{0}\right)z_{k_1}^{(f_1)}\left(n_{1}\right)\right]=0,\label{eq:statistical_SM}\end{equation}
$\forall f_0,f_1=0,1, \, n_{0},n_{1}=-\nu,\cdots,N+N_W-1, \, k_0, k_1 \in \mathbb{\mathcal{\mathcal{I}}}$.
Furthermore, based on (\ref{eq:received_sample}), (\ref{eq:corr}),
and (\ref{eq:statistical_SM}), it can be obtained that the time-varying cross-correlation function of the SM-OFDM signals is zero, i.e.,
\begin{equation}
c^{\textrm{(SM)}}(m,\tau)=0, \quad \forall m, \tau.
\label{Cor_SM}\end{equation}
Consequently, from (\ref{eq:CCF}) and (\ref{Cor_SM}), it can be seen that
\begin{equation}
C^{\textrm{(SM)}}(\alpha,\tau)=0, \quad \forall \alpha, \tau.
\label{CCF_SM}\end{equation}
\begin{flushleft}
\textbf{AL-OFDM}
\par\end{flushleft}
By using (\ref{eq:AL}), (\ref{IFFT}), (\ref{eq:Win}), the complex conjugation property of the Fourier transform, and following \cite{Mar_Dob}, one can obtain
\begin{equation*}
\mathrm{E}\left[z_{2k_0+u_0}^{(f_0)}\left(n_{0}\right)z_{2k_1+u_1}^{(f_1)}\left(n_{1}\right)\right]= 
\end{equation*}
\begin{equation}
\left\{ \begin{array}{ll}
\sigma_{s}^{2}W_{n_0}W_{n_1}\delta\left(mod\left(n_{0}+n_{1},N\right)\right)\delta\left(k_0-k_1\right) & \forall\left(f_0=0,\, f_1=1,\, u_0=0,\, u_1=1\right),\\
 & \,\,\,\,\left(f_0=1,\, f_1=0,\, u_0=1,\, u_1=0\right),\\
-\sigma_{s}^{2}W_{n_0}W_{n_1}\delta\left(mod\left(n_{0}+n_{1},N\right)\right)\delta\left(k_0-k_1\right) & \forall\left(f_0=1,\, f_1=0,\: u_0=0,\, u_1=1\right),\\
 & \,\,\,\,\left(f_0=0,\, f_1=1,\, u_0=1,\, u_1=0\right),\\
0 & \textrm{otherwise}.\end{array}\right.\label{Cor_z}\end{equation}
In other words, this correlation is non-zero for adjacent OFDM symbols within the same AL block (due to the structure of the AL coding matrix), and for samples within such OFDM symbols which satisfy the condition $mod\left(n_0+n_1,N\right)=0$.
By using (\ref{eq:received_sample}) and (\ref{eq:corr}), the time-varying cross-correlation function of the received AL-OFDM signal is expressed as (the proof is provided in Appendix A)
\begin{eqnarray}
&c^{\textrm{(AL)}}(m,\tau)=\sum\limits_{p_0,p_1=0}^{L_p-1}(h_{00}(p_0)h_{11}(p_1)-h_{01}(p_0)h_{10}(p_1))\sum\limits_{k=-\infty}^{+\infty} \delta\left(m-2k(N+\nu)-\vartheta(p_0)\right)  \otimes &\nonumber \\& \sum\limits_{n_0,n_1=-\nu}^{N+N_W-1}\sigma_{s}^{2}W_{n_0}W_{n_1}\delta\left(mod\left(n_{0}+n_{1},N\right)\right) (\delta(m-n_0-\vartheta(p_0))& \label{Cor_AL}\\&\delta(\tau-(N+\nu)+n_0-n_1+\vartheta(p_0)-\vartheta(p_1))& \nonumber \\& -\delta(m-(N+\nu)-n_0-\vartheta(p_0))\delta(\tau+(N+\nu)+n_0-n_1+\vartheta(p_0)-\vartheta(p_1))),& \nonumber
\end{eqnarray}
\noindent where $\otimes$ is the convolution operator. From (\ref{Cor_AL}), one can see that the time-varying cross-correlation function \linebreak $c^{\textrm{(AL)}}(m,\tau)$ is periodic in $m$ with the fundamental period  $M_0=2(N+\nu)$, which proves that the AL-OFDM signal exhibits second-order cyclostationarity. Furthermore, also according to the results in Appendix A, the time-varying cross-correlation function of the received AL-OFDM signal for the special case of flat fading channel can be easily obtained as
\begin{eqnarray}
&c^{\textrm{(AL)}}(m,\tau)=(h_{00}h_{11}-h_{01}h_{10})\sum\limits_{k=-\infty}^{+\infty} \delta\left(m-2k(N+\nu)\right) \otimes \sum\limits_{n_0,n_1=-\nu}^{N+N_W-1}\sigma_{s}^{2}W_{n_0}W_{n_1}& \nonumber \\& \delta\left(mod\left(n_{0}+n_{1},N\right)\right) (\delta(m-n_0)\delta(\tau-(N+\nu)+n_0-n_1)& \label{Cor_rec2} \\&-\delta(m-(N+\nu)-n_0)\delta(\tau+(N+\nu)+n_0-n_1)).&\nonumber\end{eqnarray}
By calculating the Fourier coefficients of $c^{\textrm{(AL)}}(m,\tau)$ in (\ref{Cor_rec2}), one can easily show that the CCF of the received AL-OFDM signal affected by flat fading channel is expressed as
\begin{eqnarray}
&C^{\textrm{(AL)}}(\alpha,\tau)=\frac{h_{00}h_{11}-h_{01}h_{10}}{2(N+\nu)}\sigma_{s}^{2}\sum\limits_{n_0,n_1=-\nu}^{N+N_W-1}W_{n_0}W_{n_1}\delta\left(mod\left(n_{0}+n_{1},N\right)\right)& \label{CCF_flat} \\&  (\delta(\tau-(N+\nu)+n_0-n_1)e^{-j2\pi \alpha n_0}-\delta(\tau+(N+\nu)+n_0-n_1)e^{-j2\pi \alpha (N+\nu+n_0)}),&\nonumber\end{eqnarray}
where the corresponding CFs are given as $\alpha=\frac{\ell}{2(N+\nu)}, \, \ell \in \mathbb{\mathcal{\mathcal{I}}}$.
By considering the conditions imposed by the three Kronecker delta functions on the right-hand side of (\ref{CCF_flat}) and after some mathematical manipulations, the CCF can be further expressed as (the proof is provided in Appendix B)
\begin{equation}
C^{\textrm{(AL)}}(\alpha,\tau)=\left\{\begin{array}{ll} sgn(\tau) g_1(\tau) e^{-j\pi \alpha (2N+\nu-\tau)}, & |\tau| \in \mathbb{\mathcal{\mathcal{I}}}_1 \cap \mathbb{\mathcal{\mathcal{I}}}_0^c,\\
sgn(\tau) \sum\limits_{q=0}^{1}g_q(\tau) e^{-j\pi \alpha ((q+1)N+\nu-\tau)}, & |\tau| \in \mathbb{\mathcal{\mathcal{I}}}_0 \cap \mathbb{\mathcal{\mathcal{I}}}_2^c,\\
sgn(\tau) \sum\limits_{q=0}^{2}g_q(\tau) e^{-j\pi \alpha ((q+1)N+\nu-\tau)}, & |\tau| \in \mathbb{\mathcal{\mathcal{I}}}_2,\\
0, & otherwise,
\end{array}\right. \label{CCF_flat2}
\end{equation}
where $sgn(.)$ is the signum function, $|.|$ is the absolute value, $\cap$ represents the intersection operator, the superscript $c$ denotes the set complement, $g_q(\tau)=\frac{h_{00}h_{11}-h_{01}h_{10}}{2(N+\nu)}\sigma_{s}^{2}W_{\frac{qN-|\tau|+N+\nu}{2}}W_{\frac{qN+|\tau|-N-\nu}{2}}$, and $\mathbb{\mathcal{\mathcal{I}}}_0$, $\mathbb{\mathcal{\mathcal{I}}}_1$, and $\mathbb{\mathcal{\mathcal{I}}}_2$ are defined as
\begin{equation}
\mathbb{\mathcal{\mathcal{I}}}_0=\{N-\nu,N-\nu+2,...,N+3\nu-2,N+3\nu\}, \label{eq:I0}
\end{equation}
\begin{equation}
\mathbb{\mathcal{\mathcal{I}}}_1=\{\nu-2N_W+2,\nu-2N_W+4,...,2N+\nu+2N_W-4,2N+\nu+2N_W-2\}, \label{eq:I1}
\end{equation}
\begin{equation}
\mathbb{\mathcal{\mathcal{I}}}_2=\{N+\nu-2N_W+2,N+\nu-2N_W+4,...,N+\nu+2N_W-4,N+\nu+2N_W-2\}. \label{eq:I2}
\end{equation}
From (\ref{CCF_flat2}), one can notice that there are three regions of $\tau$ for which the CCF is non-zero; in these regions, CCF consists of one term (when $|\tau| \in \mathbb{\mathcal{\mathcal{I}}}_1 \cap \mathbb{\mathcal{\mathcal{I}}}_0^c$), two terms (when $|\tau| \in \mathbb{\mathcal{\mathcal{I}}}_0 \cap \mathbb{\mathcal{\mathcal{I}}}_2^c$), and three terms (when $|\tau| \in \mathbb{\mathcal{\mathcal{I}}}_2$), respectively. Based on the results in (\ref{CCF_SM}) and (\ref{CCF_flat2}), it is clear that CCF represents a discriminating feature for the SM-OFDM and AL-OFDM signals; in the sequel, this will be exploited to develop a signal identification algorithm.
 
\subsection{Proposed Algorithm for SM-OFDM and AL-OFDM Signal  Identification\label{sec:Algorithm}}
\begin{figure}
 \centering
 \includegraphics[width=0.7\linewidth]{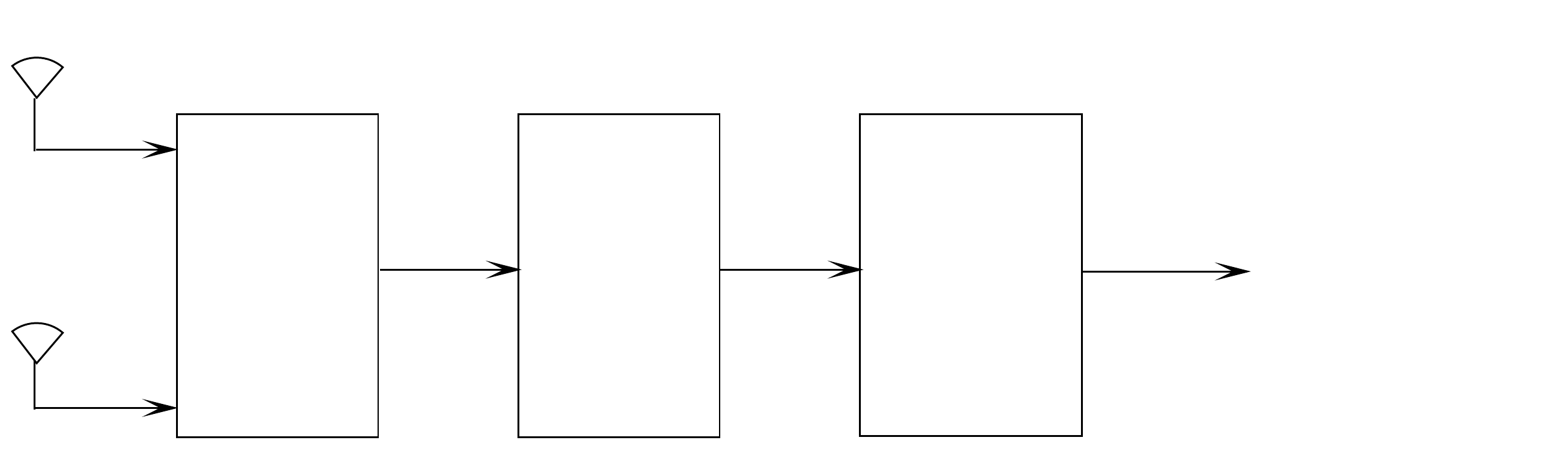}
 \caption{Block diagram of the proposed identification algorithm ($N_r=2$).\label{fig:rec}}
 \end{figure}

\textit{Two Receive Antennas ($N_r=2$) Case:} The block diagram of the proposed identification algorithm is presented in Fig. \ref{fig:rec}. In the first step, the CCF is estimated at CFs $\alpha=0, \alpha_0, -\alpha_0$, with $\alpha_0=\frac{1}{2(N+\nu)}$, and different values of $\tau$ for which CCF is non-zero for the AL-OFDM signals (the details on the delay values are provided later on in the paper). The estimate of the  CCF at CF $\alpha$ and delay $\tau$ is \cite{Test_cyc1}
\begin{equation}
\hat{C}(\alpha,\tau)=\frac{1}{M_r}\sum\limits_{m=0}^{M_r-1}r^{(0)}(m)r^{(1)}(m+\tau)e^{-j2\pi\alpha m}, \label{est_CCF}
\end{equation}
\noindent where $M_r$ is the number of received samples, equal to $N_s(N+\nu)$, with $N_s$ as the number of OFDM symbols. 
\par
In the second step, the estimated CCF magnitude is compared with a threshold set up based on  a constant false alarm criterion. The probability of false alarm is defined as the probability of identifying the received signal as AL-OFDM while it is SM-OFDM. An analytical closed form expression of the false alarm probability is obtained based on the distribution of the CCF magnitude estimate for the SM-OFDM signals. According to \cite{Test_cyc1}, the CCF estimate has an asymptotic complex Gaussian distribution. Consequently, based on (\ref{CCF_SM}), one can further infer that the CCF magnitude estimate of the SM-OFDM signal has an asymptotic Rayleigh distribution. Hence, if the CCF for a single CF $\alpha$ and delay $\tau$ is used as a discriminating feature, the probability of false alarm is calculated using the complementary cumulative density function of the Rayleigh distribution as
  \begin{equation}
P_f=\exp(-\frac{\Gamma^2}{\sigma^2}), \label{eq:Pf}
\end{equation}
\noindent where $\Gamma$ is the threshold and $\sigma^2$ is the variance of the CCF magnitude estimate for the SM-OFDM signal\nolinebreak \footnote{Note that the variance $\sigma^2$ can be estimated based on the CCF magnitude estimate of the received signal at any arbitrary CF and delays $\tau > 2(N+\nu)$. In such cases, the distribution of the CCF magnitude estimate is the same regardless of the received signal type, CF, and delay.}. When the CCF at various $\alpha$ and $\tau$ values is used for identification, the $\kappa$-out-of-$\zeta$ rule is employed for decision making, i.e., if $\kappa$ out of $\zeta$ estimated CCF magnitudes exceed the threshold, the signal is identified as AL-OFDM; otherwise, it is identified as SM-OFDM (see the third block in Fig. \ref{fig:rec}, where $\lambda$ is the final decision made by the algorithm). In this case, the probability of false alarm is\footnote{Note that (\ref{eq:Pft}) is written under the assumption that the estimated CCF of the SM-OFDM signal is uncorrelated for different CFs and delays. The validity of this assumption was verified through extensive simulations.}
\begin{equation}
P_F=\sum\limits_{\ell=\kappa}^{\zeta}{\zeta \choose \ell}P_f^{\ell}(1-P_f)^{\zeta-\ell}. \label{eq:Pft}
\end{equation}
 In this case, for a constant false alarm rate $P_F$, $P_f$ is calculated using (\ref{eq:Pft}), and then the threshold value is obtained from (\ref{eq:Pf}). The remaining problem is the selection of the parameters $\kappa$ and $\zeta$. From (\ref{CCF_flat2}), one can notice that for each value of $\alpha$, the CCF has a larger magnitude when $|\tau| \in \mathbb{\mathcal{\mathcal{I}}}_0$ (the second and third branches of (\ref{CCF_flat2})). When a single CF is used, we consider $|\tau| \in \mathbb{\mathcal{\mathcal{I}}}_0 \cup \{N-\nu+1,N-\nu+3,...,N+3\nu-3,N+3\nu-1\}$\footnote{Note that the closed-form expression for the CCF at CF $\alpha$ and delay $\tau$ in (\ref{CCF_flat2}) was obtained under the assumption of flat fading channel. For the frequency-selective fading channel, in addition to the set of delays $\mathbb{\mathcal{\mathcal{I}}}_0$, we also include the set $\{N-\nu+1, N-\nu +3, …, N+3\nu-3, N+3\nu-1\}$. For the reason of additionally considering this set of delays, the reader is referred to "Analytical and simulation results for the CCF magnitude," as well as Figs. \ref{fig:32sfl} and \ref{fig:32sfs} in Section \ref{sec:sims}.}, which leads to $\zeta=8\nu+2$. Furthermore, when three CFs are used, $\zeta=24\nu+6$. After extensive simulations run under various scenarios, we selected $\kappa=\nu/2$, as providing a good performance. A summary of the proposed algorithm is provided as follows.

\floatname{algorithm}{}

\begin{algorithm}
\renewcommand{\thealgorithm}{}
\caption{\textbf{Proposed algorithm}}
\begin{algorithmic}[0]
\STATE \textbf{Input:} The received signals $r^{(v)}(m)$, $v=0, 1$, the number of subcarriers $N$, and the CPR $\nu$.
\STATE - Estimate the CCF using (\ref{est_CCF}) at CFs $\alpha=0, \alpha_0, -\alpha_0$, and for $|\tau| \in \mathbb{\mathcal{\mathcal{I}}}_0 \cup \{N-\nu+1,N-\nu+3,...,N+3\nu-3,N+3\nu-1\}$ (for the discriminating feature) and $\tau=2(N+\nu)+1,...,3(N+\nu)$ (for estimating the variance $\sigma^2$).
\STATE - Estimate the parameter $\sigma^2$ from the CCF at CFs $\alpha=0, \alpha_0, -\alpha_0$, and for $\tau=2(N+\nu)+1,...,3(N+\nu)$.
\STATE - Calculate the threshold, $\Gamma$, using (\ref{eq:Pf}) and (\ref{eq:Pft}).
\STATE - {Compare the estimated CCF magnitude at CFs $\alpha=0, \alpha_0, -\alpha_0$, and for $|\tau| \in \mathbb{\mathcal{\mathcal{I}}}_0$ with $\Gamma$.}
\IF {at least $\kappa$ estimated CCF magnitudes exceed the threshold,}
\STATE - {the received signal is identified as AL-OFDM.}
\ELSE
\STATE - {the received signal is identified as SM-OFDM.}
\ENDIF
\end{algorithmic}
\end{algorithm}
\par
\textit{Computational complexity:} We evaluate the computational complexity of the proposed identification algorithm through the number of floating point operations (flops) \cite{matrixcomp}, where a complex multiplication and addition require six and two flops, respectively. According to the algorithm description, CCF is estimated for $3(N+9\nu+2)$ combinations of CFs and delays. Based on (\ref{est_CCF}), one can easily see that the number of complex multiplications and additions required to calculate the CCF at a certain CF and delay equals to $2N_s(N+\nu)$ and $N_s(N+\nu)-1$, respectively. By considering that the thresholding does not require additional complexity, it is straightforward that the number of flops needed by the algorithm equals to $3(14N_s(N+\nu)-2)(N+9\nu+2)$. It is worth noting that with an average Intel Core i750, the identification process takes $0.162$ sec for $N_s=2,000$, $N=64$, and $\nu=8$, whereas with an IBM Sequoia supercomputer, this time reduces to $7.5\times10^{-8}$ sec.
\par 
\textit{Number of Receive Antennas $N_r>2$ Case:} Here we extend the proposed algorithm to the case of  $N_r>2$. For each pair of receive antennas, $(i_0,i_1)$, $i_0<i_1$, $i_0,i_1=0,1,...,N_r-1$, we define the corresponding CCF. Consequently, for each pair of receive antennas, the CCF is estimated using (\ref{est_CCF}) at the CFs and delays considered for the case of $N_r=2$. Then, with the values of $\zeta$ and $\kappa$ in (\ref{eq:Pft}) scaled by $\frac{N_r(N_r-1)}{2}$, the same steps are applied as for the case of $N_r=2$. Note that $\frac{N_r(N_r-1)}{2}$ represents the number of different combinations of two received antennas. It is easy to notice that the complexity of the proposed  algorithm for the $N_r>2$ case is also scaled by $\frac{N_r(N_r-1)}{2}$. 
\section{Simulation Results}\label{sec:sims}
In this section, we compare the analytical and simulation results for the CCF magnitude, investigate the performance of the proposed algorithm, and compare it with that of the algorithm in \cite{Mar_Dob}. 
\par
\textit{Simulation setup:} Unless otherwise indicated, we consider an OFDM signal with quadrature phase-shift-keying (QPSK) modulation, $N=64$, $\nu=8$, a raised-cosine window with $N_W=2$, and $N_r=2$. The carrier frequency is $f_c=2.5$ GHz and the duration of the OFDM symbol is $T=91.4$ $\mu$sec. The probability of false alarm is $P_F=10^{-2}$, and the number of OFDM symbols is $N_s=2,$$000$. The received signal is affected by AWGN with variance $\sigma_w^2$ and a frequency-selective fading channel consisting of $L_{p}=4$ statistically independent taps, each being a zero-mean complex Gaussian random variable. The channel is characterized by an exponential
power delay profile, $\sigma^{2}(p)=B_{h}exp(-p/5),$ where $p=0,...,L_p-1$
and $B_{h}$ is chosen such that the average power is normalized to unity \cite{Mar_Dob}. The SNR is defined as $10 \log_{10}(\frac{2}{\sigma_w^2})$. The probability of correct identification, $P(\lambda=\xi|\xi)$, with $\lambda$ as the decided signal type and $\xi=\mbox{SM},\mbox{AL}$, is considered as a performance measure and is evaluated using Monte Carlo simulations with $1,$$000$ trials. 
\par
\begin{figure}
	\centering
		\includegraphics[width=1\linewidth]{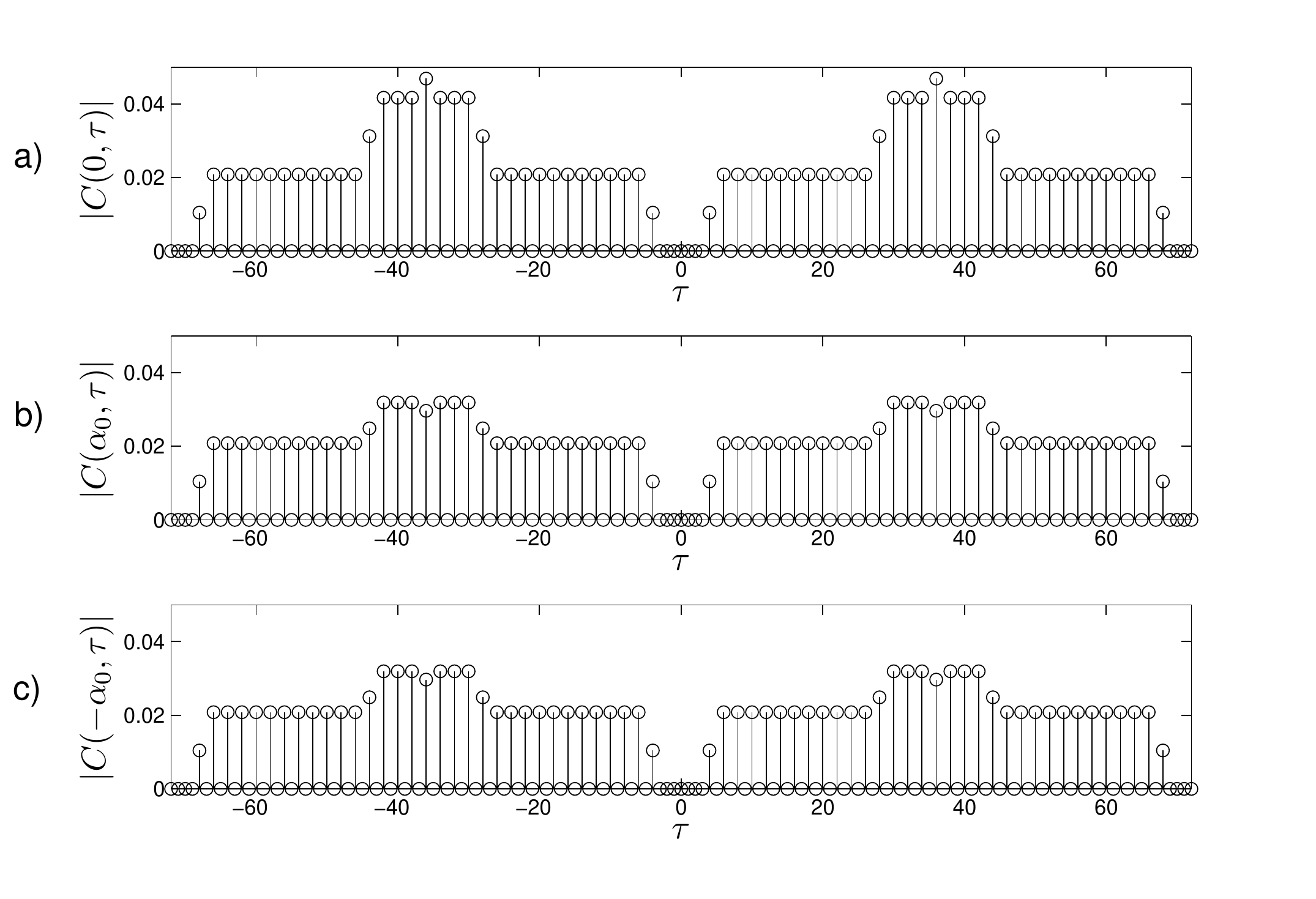}
	\caption{CCF magnitude calculated from (\ref{CCF_flat2}) versus delay $\tau$ at CFs a) zero b) $\alpha_0=\frac{1}{2(N+\nu)}$ c) $-\alpha_0=-\frac{1}{2(N+\nu)}$ for flat Rayleigh fading channel.}
	\label{fig:32a}
\end{figure}
\begin{figure}
	\centering
		\includegraphics[width=1\linewidth]{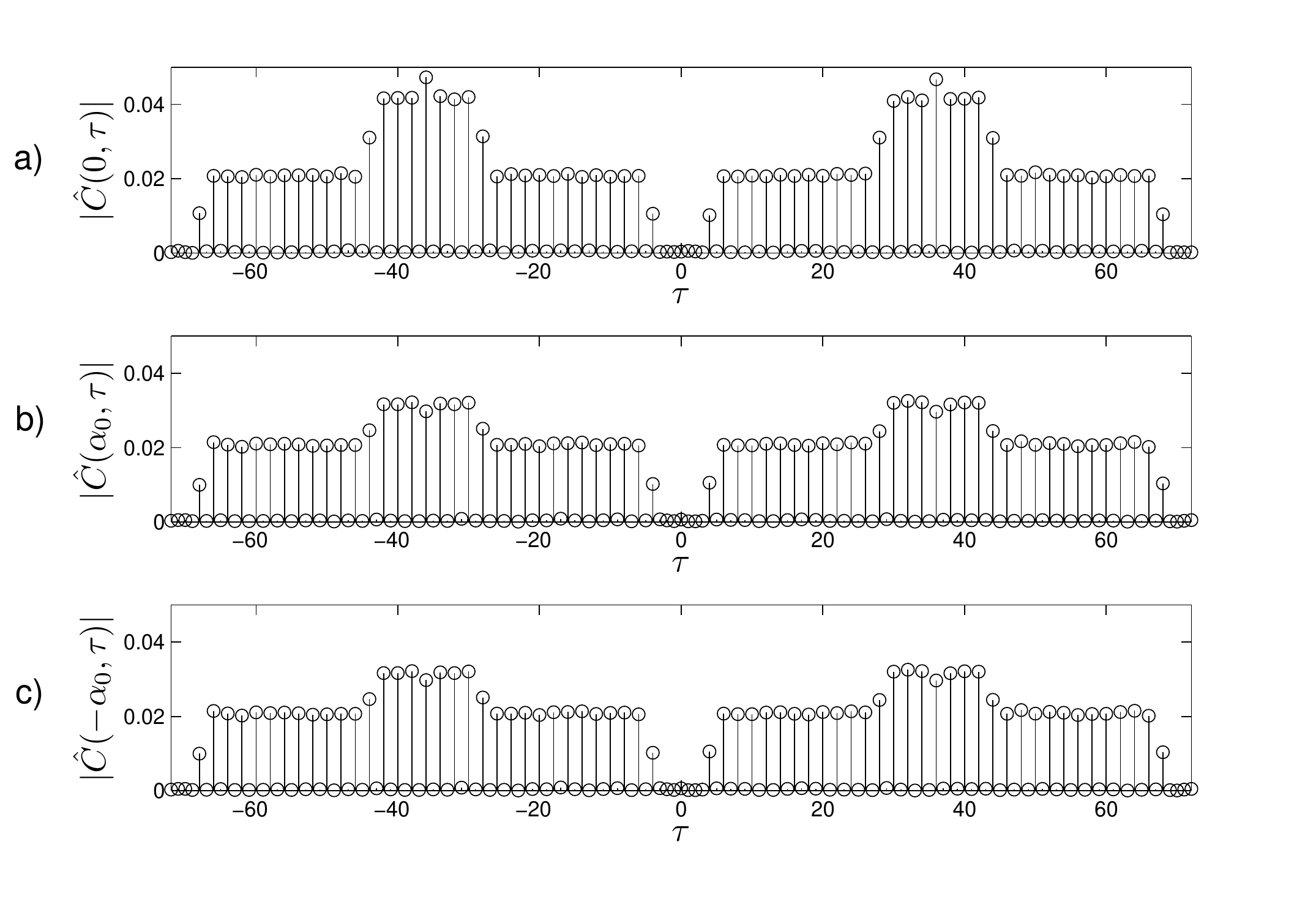}
		\caption{CCF magnitude estimate versus delay $\tau$ at CFs a) zero b) $\alpha_0=\frac{1}{2(N+\nu)}$ c) $-\alpha_0=-\frac{1}{2(N+\nu)}$ for flat Rayleigh fading channel.}
	\label{fig:32sfl}
\end{figure}
\begin{figure}
	\centering
		\includegraphics[width=1\linewidth]{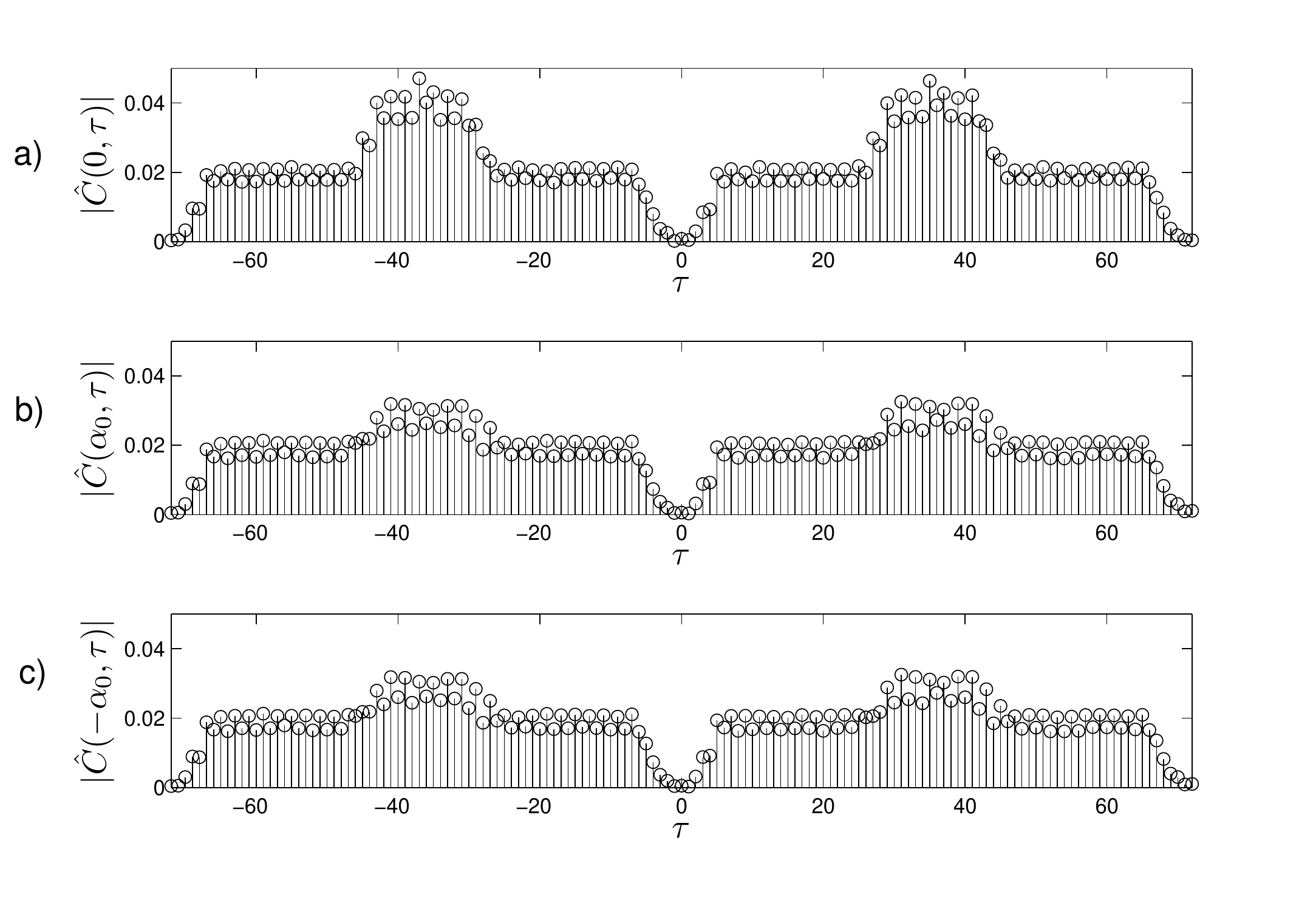}
		\caption{CCF magnitude estimate versus delay $\tau$ at CFs a) zero b) $\alpha_0=\frac{1}{2(N+\nu)}$ c) $-\alpha_0=-\frac{1}{2(N+\nu)}$ for frequency-selective channel.}
	\label{fig:32sfs}
\end{figure}

 \textit{Analytical and simulation results for the CCF magnitude:} Figs. \ref{fig:32a} and \ref{fig:32sfl} present analytical and simulation results for the CCF magnitude versus delay $\tau$, for $|\tau|=0,1,...,2N+2\nu-2,2N+2\nu-1$ and at CFs $\alpha=0,\alpha_0, -\alpha_0$, with $\alpha_0=\frac{1}{2(N+\nu)}$. An OFDM signal with $N=32$, $\nu=4$, $N_W=1$, and flat Rayleigh fading channel are considered. For simulation results, SNR=10 dB and $N_s=10^6$ OFDM symbols. As can be noticed, the analytical and simulation results are in agreement, and as expected, a larger CCF magnitude is observed for $|\tau|=N-\nu,N-\nu+2,...,N+3\nu-2,N+3\nu$ (corresponding to the second and third branches on the right-hand side of (\ref{CCF_flat2})). Furthermore, Fig. \ref{fig:32sfs} presents the CCF magnitude estimate versus delay $\tau$ for the frequency-selective Rayleigh fading channel. One can see that, when compared to the flat fading channel, there is a dispersion in the CCF magnitude which results in non-zero values for odd delays. Hence, for identification purposes, we considered the CCF magnitudes for the delay range $|\tau|=N-\nu,N-\nu+1,...,N+3\nu-1,N+3\nu$ as discriminating signal features (see the formal description of the algorithm in Section III).
\par
\begin{figure}
	\centering
		\includegraphics[width=0.9\linewidth]{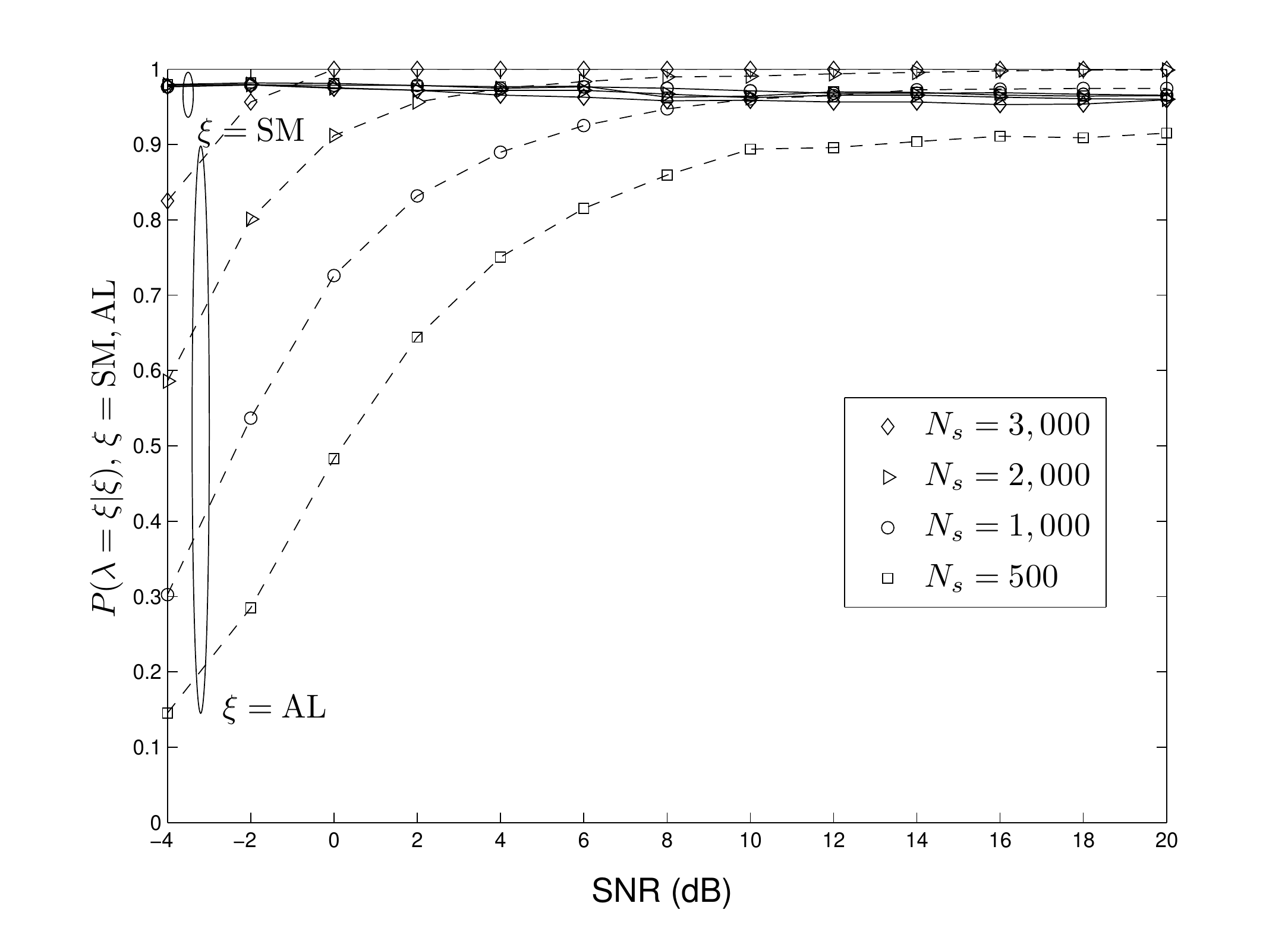}
	\caption{Probability of correct identification, $P(\lambda=\xi|\xi)$, $\xi=\mbox{SM},\mbox{AL}$, versus SNR for different numbers of received symbols, $N_s$. Solid lines are used for $\xi=\mbox{SM}$ and dashed lines for $\xi=\mbox{AL}$.}
	\label{fig:Perf1}
\end{figure}
\begin{figure}
	\centering
		\includegraphics[width=0.9\linewidth]{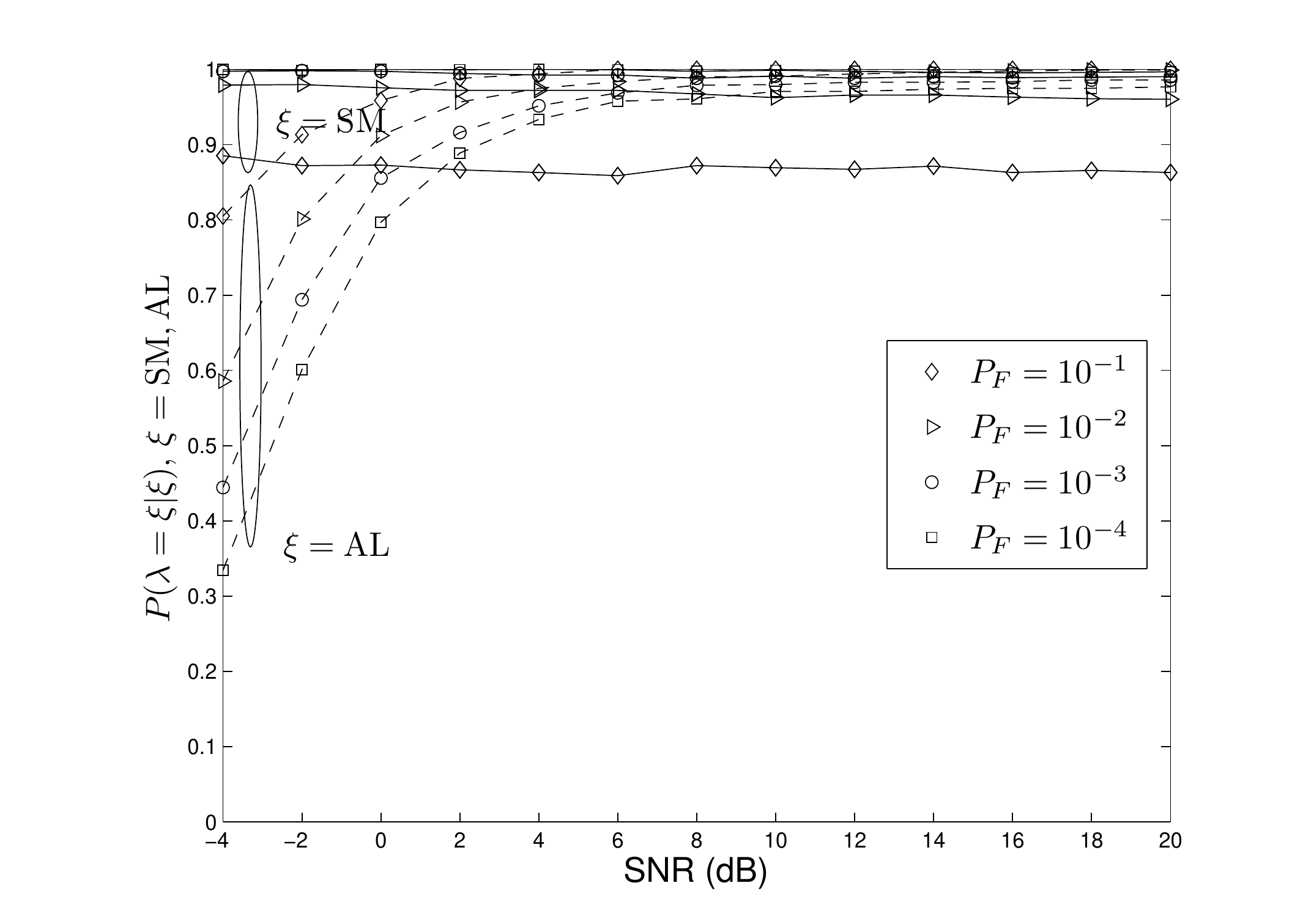}
	\caption{Probability of correct identification, $P(\lambda=\xi|\xi)$, $\xi=\mbox{SM},\mbox{AL}$, versus SNR for different values of $P_F$. Solid lines are used for $\xi=\mbox{SM}$ and dashed lines for $\xi=\mbox{AL}$.}
	\label{fig:Perf2}
\end{figure}
\begin{figure}
	\centering
		\includegraphics[width=0.9\linewidth]{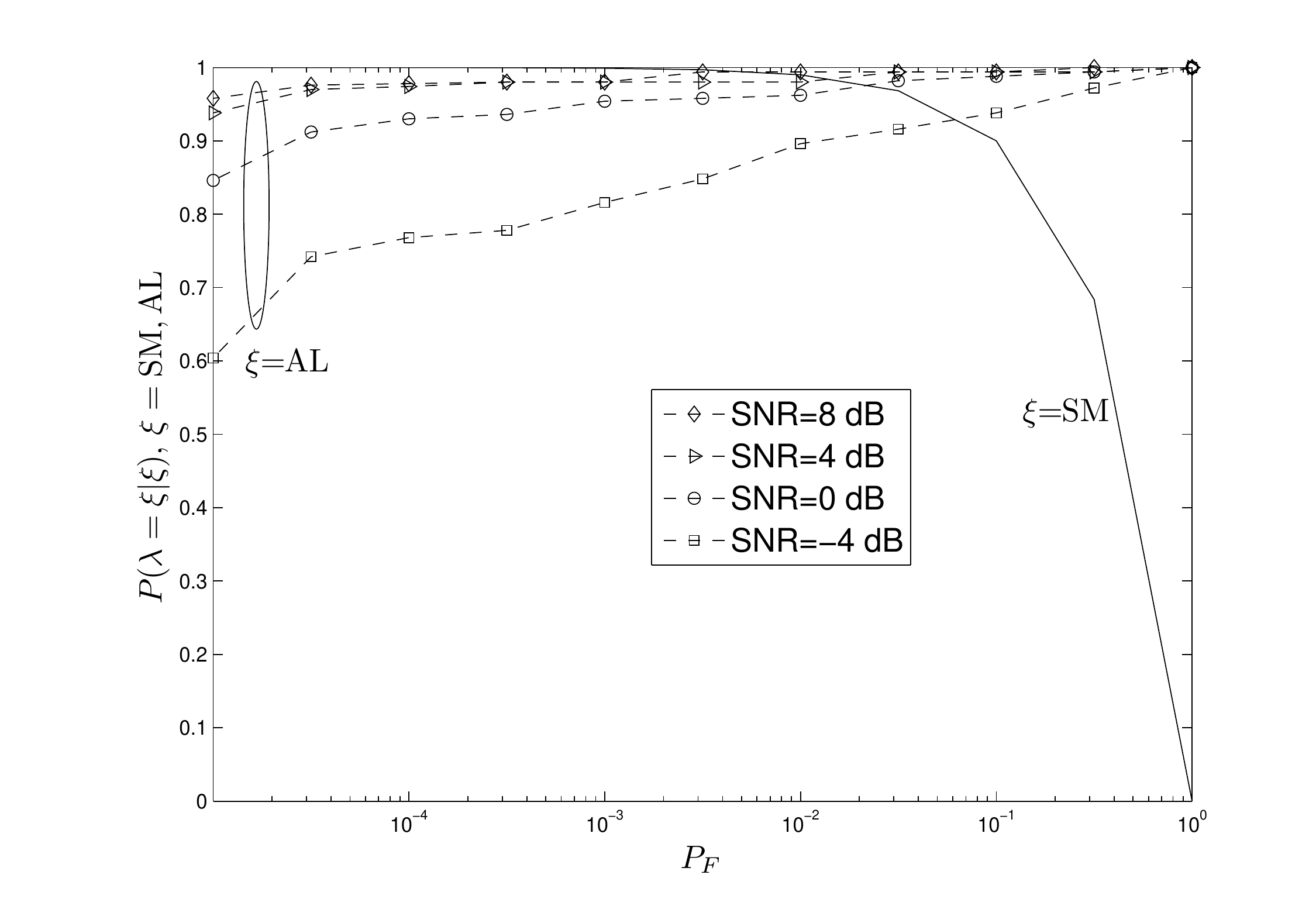}
	\caption{Probability of correct identification, $P(\lambda=\xi|\xi)$, $\xi=\mbox{SM},\mbox{AL}$, versus $P_F$ for different values of SNR. Solid line is used for $\xi=\mbox{SM}$ and dashed lines for $\xi=\mbox{AL}$.}
	\label{fig:PF}
\end{figure}

 \textit{Identification performance of the proposed algorithm:} Figs. \ref{fig:Perf1} and \ref{fig:Perf2} show the probability of correct identification, $P(\lambda=\xi|\xi)$, $\xi=\mbox{SM},\mbox{AL}$, versus SNR for different numbers of OFDM symbols, $N_s$, and probability of false alarm, $P_F$, respectively. As expected, results obtained for $P(\lambda=\mbox{SM}|\mbox{SM})$ are close to $1-P_F$ regardless of the SNR and $N_s$. $P(\lambda=\mbox{AL}|\mbox{AL})$ improves as SNR and $N_s$ increase and $P_F$ decreases. This can be easily explained, as the accuracy of the CCF magnitude estimate enhances when a larger SNR and observation period are available, and a lower threshold corresponds to a reduced $P_F$. According to Fig. \ref{fig:Perf1}, $P(\lambda=\mbox{AL}|\mbox{AL})$ approaches to one at 0 dB SNR with $N_s=3,$$000$, while 8 dB SNR is required for $N_s=2,$$000$ when $P_F=10^{-2}$. On the other hand, only 2 dB SNR is required to reach such a performance for $N_s=2,$$000$ when $P_F=10^{-1}$, as shown in Fig. \ref{fig:Perf2}. Additionally, the behaviour of $P(\lambda=\xi|\xi)$, $\xi=\mbox{SM},\mbox{AL}$, as a function of $P_F$ is provided in Fig. \ref{fig:PF}. Note that since $P(\lambda=\mbox{SM}|\mbox{SM})=1-P_F$ is the same for different values of SNR, in Fig. \ref{fig:PF}, one curve with solid line is used to show $P(\lambda=\mbox{SM}|\mbox{SM})$.
\par
\begin{figure}
	\centering
		\includegraphics[width=0.9\linewidth]{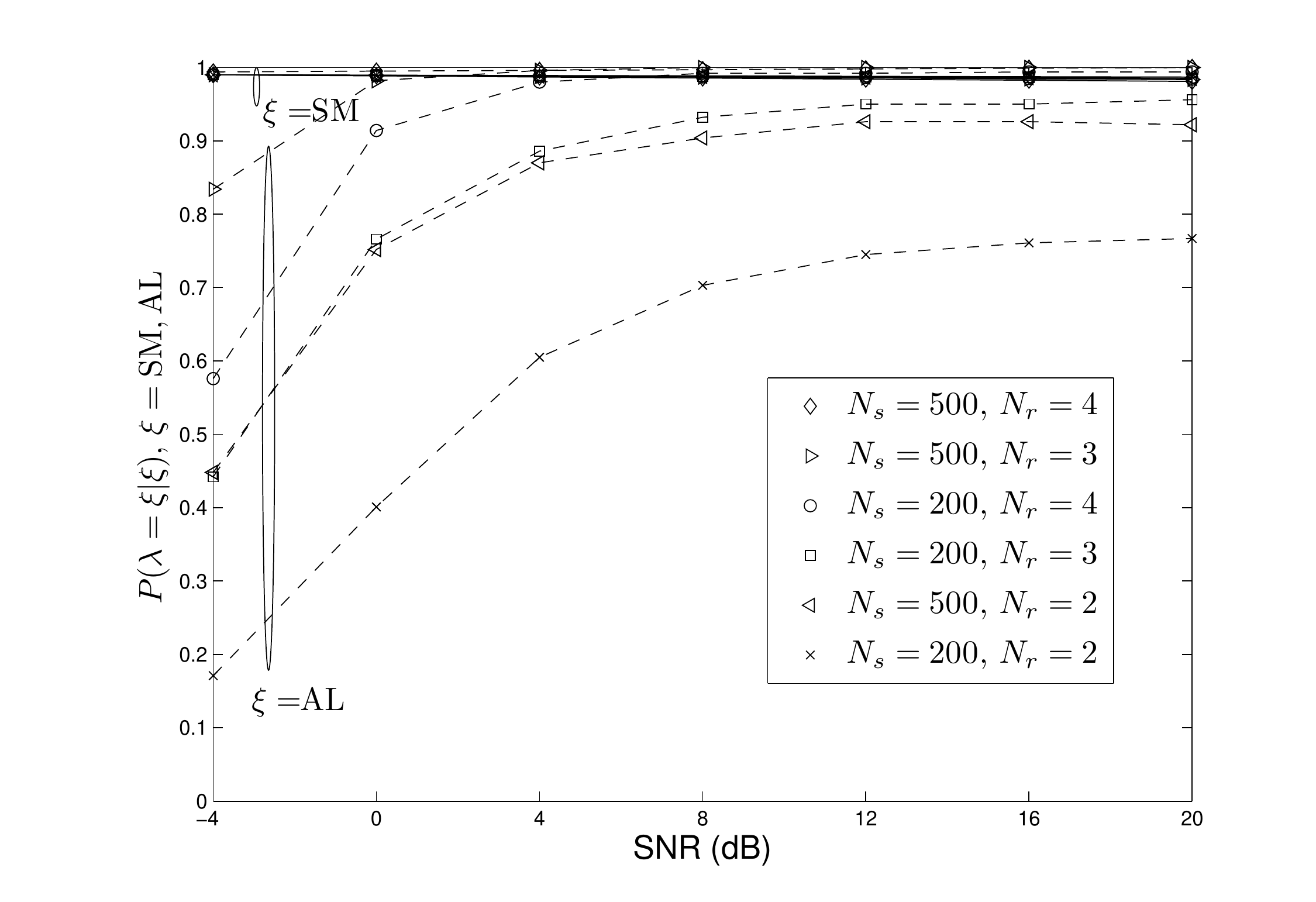}
	\caption{Probability of correct identification, $P(\lambda=\xi|\xi)$, $\xi=\mbox{SM},\mbox{AL}$, versus SNR for different values of $N_r$ and $N_s$. Solid line is used for $\xi=\mbox{SM}$ and dashed lines for $\xi=\mbox{AL}$.}
	\label{fig:Nr}
\end{figure}

Fig. \ref{fig:Nr} shows the probability of correct identification, $P(\lambda=\xi|\xi)$, $\xi=\mbox{SM},\mbox{AL}$, versus SNR for different numbers of OFDM symbols, $N_s$, and receive antenna, $N_r$. According to Fig. \ref{fig:Nr}, one can see that by increasing the number of receive antennas, the performance improves. As such, a certain probability of correct identification is achieved at lower SNR and/or with shorter observation time. For example, $P(\lambda=\mbox{AL}|\mbox{AL})=0.9$ is achieved with $N_s=500$ at SNR = 5.6 dB and -1.8 dB for $N_r=2$ and $N_r=3$, respectively. In other words, by increasing the number of receive antennas from two to three, a 7.4 dB performance gain is obtained. From Fig. \ref{fig:Nr}, one can further see that with $N_s=200$ and $N_r=2$, $P(\lambda=\mbox{AL}|\mbox{AL})$ does not reach $0.9$, whereas $P(\lambda=\mbox{AL}|\mbox{AL})=0.9$ is obtained at SNR= 5.3 dB and -0.2 dB for $N_r=3$ and $N_r=4$, respectively.
 \par
\begin{figure}
	\centering
		\includegraphics[width=0.9\linewidth]{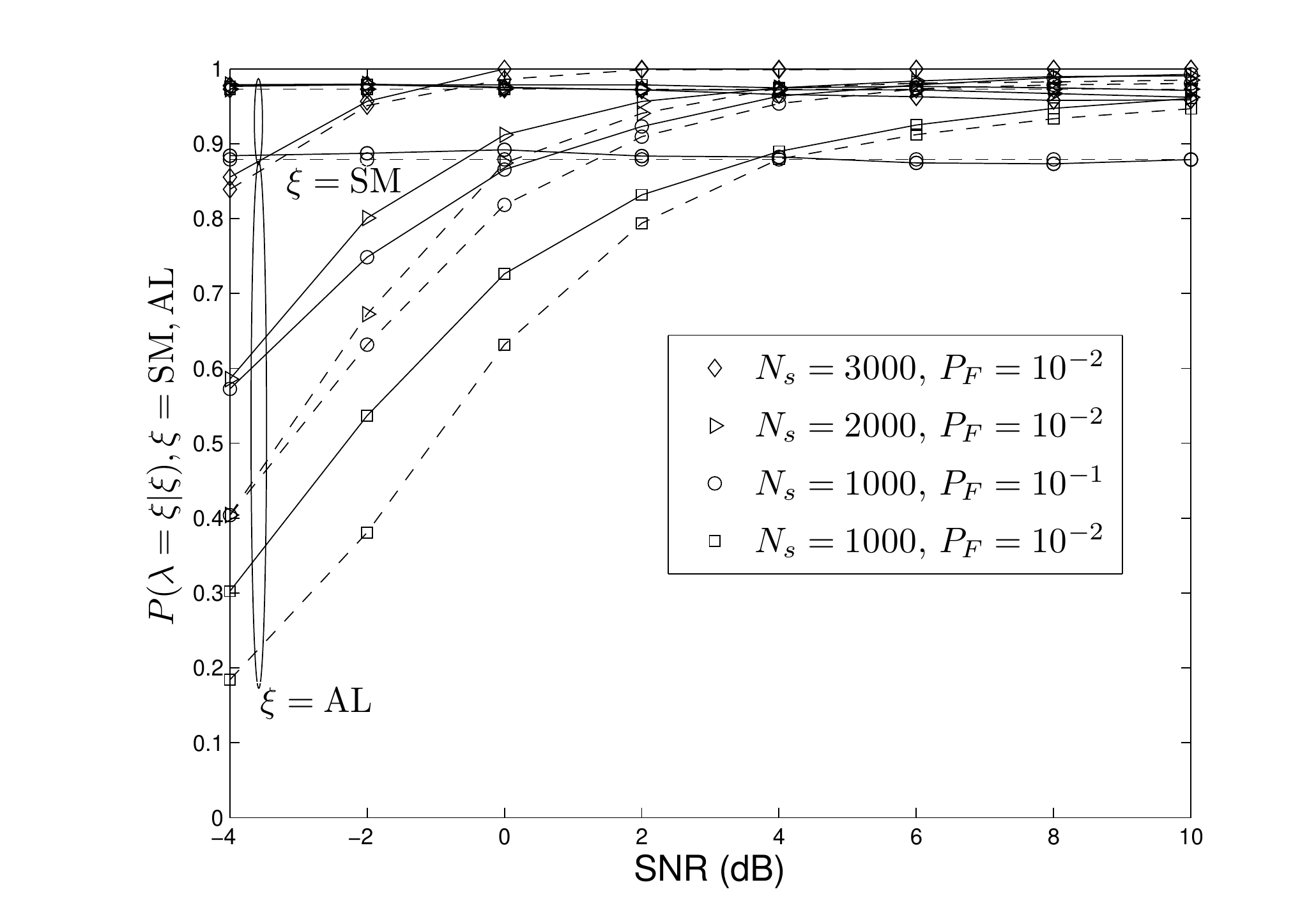}
	\caption{Performance comparison with the algorithm in \cite{Mar_Dob} for different $N_s$ and $P_F$ values. Solid lines are used for the proposed algorithm and dashed lines for the one in \cite{Mar_Dob}.}
	\label{fig:Compare1}
\end{figure}

 \textit{Performance comparison with the algorithm in \cite{Mar_Dob}:} Fig. \ref{fig:Compare1} shows a performance comparison with the algorithm in \cite{Mar_Dob}, for different values of $N_s$ and $P_F$. One can observe that while both algorithms provide a similar performance in terms of $P(\lambda=\mbox{SM}|\mbox{SM})$, the proposed algorithm outperforms the one in \cite{Mar_Dob} for $P(\lambda=\mbox{AL}|\mbox{AL})$, especially at lower SNR.
\par
In the following, we investigate the robustness of the proposed algorithm and the one in \cite{Mar_Dob} to diverse impairments, i.e., phase noise, frequency offset, timing offset, and channel conditions.\par
\begin{figure}
	\centering
		\includegraphics[width=0.9\linewidth]{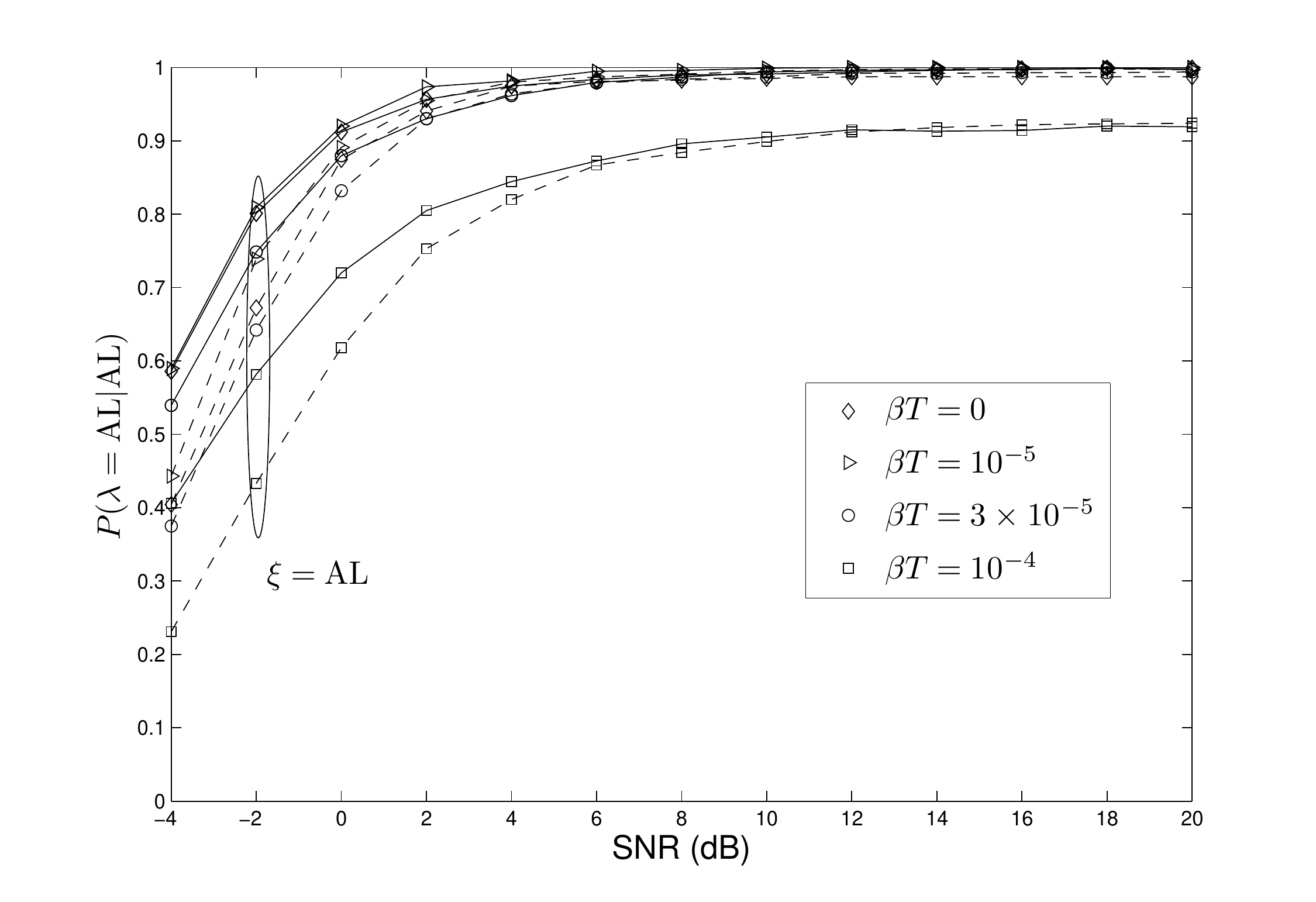}
	\caption{Sensitivity of the proposed algorithm (solid lines) and the one in \cite{Mar_Dob} (dashed lines) to phase noise.}
	\label{fig:PN1}
\end{figure}

 \textit{Effect of phase noise:} The phase noise is modeled as a Wiener process with rate $\beta T$, where $\beta$ is the two-sided 3 dB bandwidth of the Lorenzian distribution spectrum \cite{Phasenoise1}. Fig. \ref{fig:PN1} shows the probability of correct identification, $P(\lambda=\mbox{AL}|\mbox{AL})$\footnote{As the impairments in the received signal do not affect the results for the SM-OFDM signal identification, we only show the effect of the signal impairments on the probability of correct identification of AL-OFDM signals.}, versus SNR for different values of the phase noise rate.
 As can be seen, the proposed algorithm is relatively robust for $\beta T < 3\times 10^{-5}$, and its performance starts degrading for $\beta T=10^{-4}$. This can be explained based on the results obtained in Appendix B for the dependency of the CC magnitude estimate on the phase noise. According to these results, for $N_s=2,$$000$, the CCF magnitude estimate is scaled with a factor of $0.9692$, $0.9114$, and $0.7426$ for $\beta T=10^{-5}$, $3\times 10^{-5}$, and $10^{-4}$, respectively. Clearly, a reduction in the CCF magnitude leads to a performance degradation. Additionally, from Fig. \ref{fig:PN1}, one can see that the proposed algorithm is more robust to the phase noise when compared with the algorithm in \cite{Mar_Dob}.
\par
\begin{figure}
	\centering
		\includegraphics[width=0.9\linewidth]{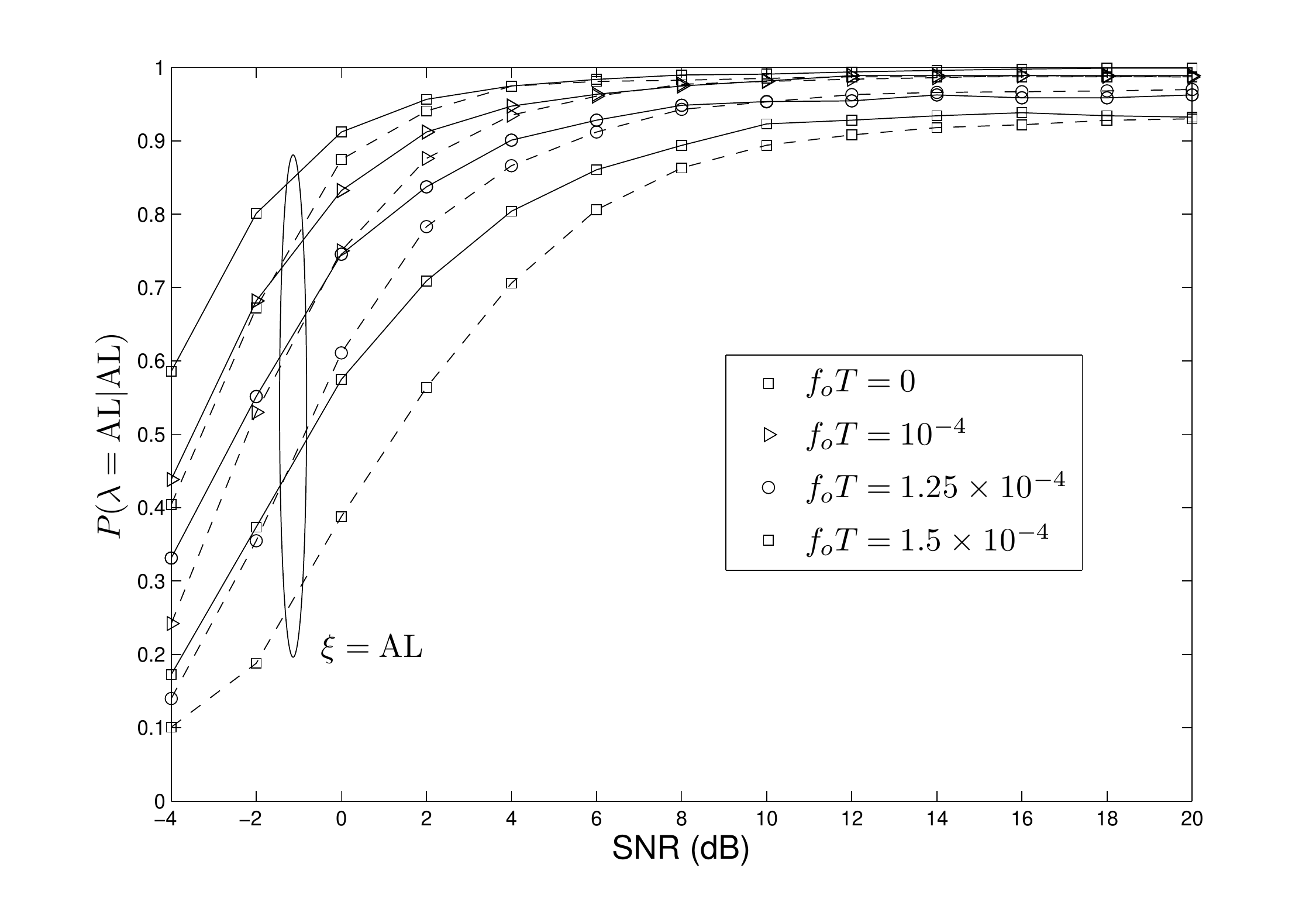}
	\caption{Sensitivity of the proposed algorithm (solid lines) and the one in \cite{Mar_Dob} (dashed lines) to frequency offset.}
	\label{fig:freq_offset}
\end{figure}

\textit{Effect of frequency offset:} Fig. \ref{fig:freq_offset} shows the probability of correct identification, $P(\lambda=\mbox{AL}|\mbox{AL})$, versus SNR for different values of the normalized frequency offset, $f_oT$, with $f_o$ as the frequency offset. As one can notice, both the proposed algorithm and the one in \cite{Mar_Dob} are robust for $f_oT \le 10^{-4}$, with the former exhibiting a better performance when compared with the latter.
\par
\begin{figure}
	\centering
		\includegraphics[width=0.9\linewidth]{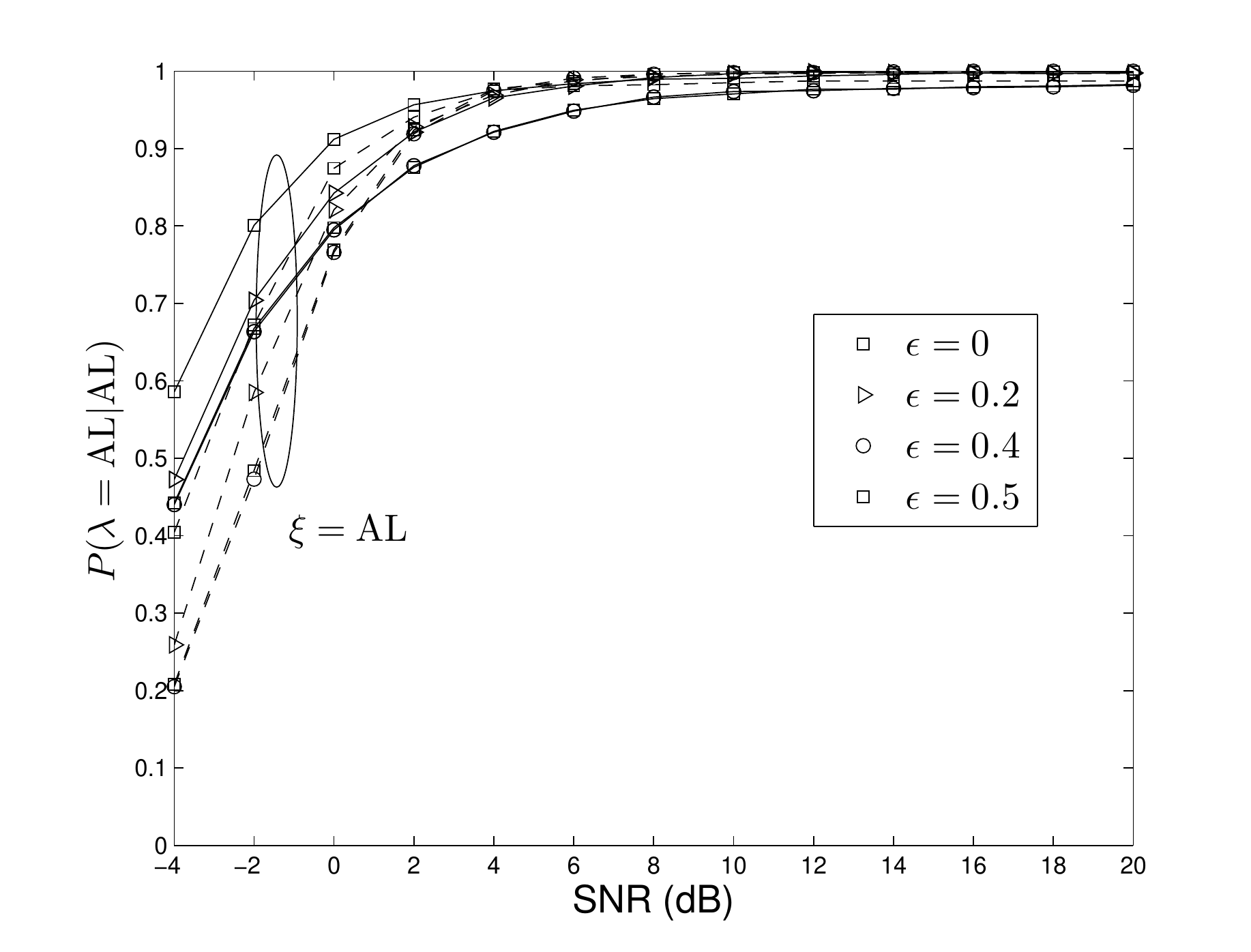}
	\caption{Sensitivity of the proposed algorithm (solid lines) and the one in \cite{Mar_Dob} (dashed lines) to timing offset.}
	\label{fig:time_offset}
\end{figure}

  
\textit{Effect of timing offset:} Fig. \ref{fig:time_offset} shows the probability of correct identification, $P(\lambda=\mbox{AL}|\mbox{AL})$, versus SNR for different values of the timing offset, $\epsilon$. By following \cite{choquese2,Yahia_TCOM}, the effect of the timing offset was obtained by passing the signal through an $\left[1-\epsilon \quad \epsilon\right]$ two path filter, when the pulse shape is rectangular. As it can be seen from Fig. \ref{fig:freq_offset}, both the proposed algorithm and the one in \cite{Mar_Dob} are relatively robust to the timing offset, with a better performance provided by the former. As expected, the performance degrades as $\epsilon$ reaches $0.5$, and in the lower SNR range.
\par
\begin{figure}
	\centering
		\includegraphics[width=0.9\linewidth]{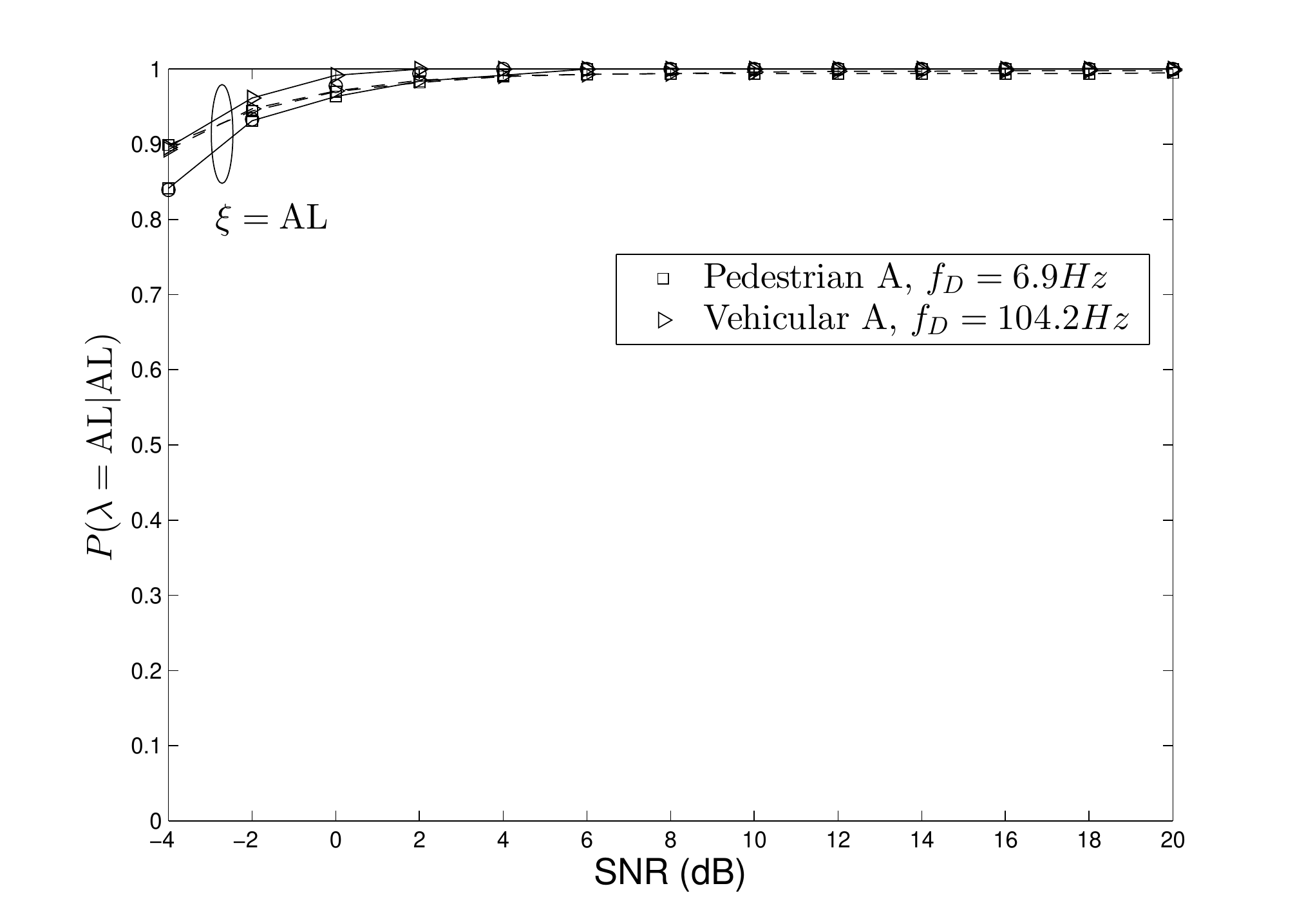}
	\caption{Performance of the proposed algorithm (solid lines) and the one in \cite{Mar_Dob} (dashed lines) for the pedestrian A and vehicular A fading channels.}
	\label{fig:channel_A}
\end{figure}

\textit{Effect of channel conditions:} We investigate the performance of the proposed algorithm and the one in \cite{Mar_Dob} in the pedestrian and vehicular A fading channels \cite{book_mo}. The maximum Doppler frequency for the pedestrian channel was $f_D=6.9$ Hz, while the maximum Doppler frequency was $f_D=104.2$ Hz for the vehicular channel. With the proposed algorithm, the channel dispersion is beneficial for identification, as
introducing additional CCF peaks (see results showed in Figs. \ref{fig:32sfl} and \ref{fig:32sfs}). As such, as can be seen from Fig. \ref{fig:channel_A}, despite a larger $f_D$, the identification performance for the vehicular A channel is slightly better than that for the pedestrian A case. Also, according to Fig. \ref{fig:channel_A}, both algorithms provide a good and similar performance under both channel conditions.

\section{Conclusions\label{sec:Conclusion}}
In this paper, we proposed a second-order cyclostationarity-based discriminating feature for SM-OFDM and AL-OFDM signals, along with a signal identification algorithm. The proposed algorithm provides a reasonable performance at relatively low SNR and with a short observation period. Furthermore, it is relatively robust to the phase noise, timing offset, and channel conditions, and outperforms the algorithm existing in the literature. As part of future work, the analysis and identification algorithm presented in this paper are planned to be extended to additional STBCs.
\section*{Acknowledgment}
The authors would like to acknowledge the constructive comments and suggestions of the Editor, Professor Jia-Chin Lin, as well as the anonymous reviewers.
\section*{Appendix A}
\renewcommand{\thesection}{\arabic{section}}
Here, we present the proof of (\ref{Cor_AL}). By replacing (\ref{eq:received_sample}) into (\ref{eq:corr}), the time-varying cross-correlation function of the received AL-OFDM signal is expressed as
\begin{eqnarray}
&c^{\textrm{(AL)}}(m,\tau)=\sum\limits_{p_0,p_1=0}^{L_p-1}h_{00}(p_0)h_{10}(p_1)\mathrm{E}\left[s^{(0)}(m-\vartheta(p_0))s^{(0)}(m+\tau-\vartheta(p_1))\right]& \nonumber\\&+h_{00}(p_0)h_{11}(p_1)\mathrm{E}\left[s^{(0)}(m-\vartheta(p_0))s^{(1)}(m+\tau-\vartheta(p_1))\right]& \nonumber\\&
+h_{01}(p_0)h_{10}(p_1)\mathrm{E}\left[s^{(1)}(m-\vartheta(p_0))s^{(0)}(m+\tau-\vartheta(p_1))\right]& \nonumber\\&
+h_{01}(p_0)h_{11}(p_1)\mathrm{E}\left[s^{(1)}(m-\vartheta(p_0))s^{(1)}(m+\tau-\vartheta(p_1))\right].&\label{Cor_rec}
\end{eqnarray}
One can easily find that the first and last terms on the right-hand side of (\ref{Cor_rec}) are zero. Additionally, by using (\ref{Tr_sig}) and (\ref{Cor_z}), it can be shown that $\mathrm{E}[s^{(1)}(m-\vartheta(p_0))s^{(0)}(m+\tau-\vartheta(p_1))]=-\mathrm{E}[s^{(0)}(m-\vartheta(p_0))$ \linebreak$\times s^{(1)}(m+\tau-\vartheta(p_1))]$. Consequently, if we define $c_s(m,\tau)=\mathrm{E}\left[s^{(0)}(m)s^{(1)}(m+\tau)\right]$ as the time-varying cross-correlation function of the transmitted signal, (\ref{Cor_rec}) can be re-written as
\begin{equation}
c^{\textrm{(AL)}}(m,\tau)=\sum\limits_{p_0,p_1=0}^{L_p-1}(h_{00}(p_0)h_{11}(p_1)-h_{01}(p_0)h_{10}(p_1))c_s^{\textrm{(AL)}}(m-\vartheta(p_0),\tau+\vartheta(p_0)-\vartheta(p_1)), \label{Cor_rec1}
\end{equation}
where 
\begin{eqnarray}
&c_s^{\textrm{(AL)}}(m,\tau)=\sum\limits_{k_0,k_1=-\infty}^{+\infty}\sum\limits_{u_0,u_1=0}^{1}\sum\limits_{n_0,n_1=-\nu}^{N+N_W-1}\mathrm{E}\left[z_{2k_0+u_0}^{(0)}\left(n_{0}\right)z_{2k_1+u_1}^{(1)}\left(n_{1}\right)\right]& \label{Cor_tr}\\&
\delta(m-(2k_0+u_0)(N+\nu)-n_0)\delta(m+\tau-(2k_1+u_1)(N+\nu)-n_1).&\nonumber\end{eqnarray}
Furthermore, based on (\ref{Cor_z}), (\ref{Cor_tr}) becomes
\begin{eqnarray}
&c_s^{\textrm{(AL)}}(m,\tau)=\sum\limits_{k=-\infty}^{+\infty}\sum\limits_{n_0,n_1=-\nu}^{N+N_W-1}\sigma_{s}^{2}W_{n_0}W_{n_1}\delta\left(mod\left(n_{0}+n_{1},N\right)\right)(\delta(m-2k(N+\nu)-n_0)&\label{Cor_tr1} \\& \delta(m+\tau-(2k+1)(N+\nu)-n_1) -\delta(m-(2k+1)(N+\nu)-n_0)\delta(m+\tau-2k(N+\nu)-n_1)),& \nonumber
\end{eqnarray}
\noindent which can be further expressed as
\begin{eqnarray}
&c_s^{\textrm{(AL)}}(m,\tau)=\sum\limits_{k=-\infty}^{+\infty} \delta\left(m-2k(N+\nu)\right) \otimes \sum\limits_{n_0,n_1=-\nu}^{N+N_W-1}\sigma_{s}^{2}W_{n_0}W_{n_1}\delta\left(mod\left(n_{0}+n_{1},N\right)\right)& \label{Cor_tr2} \\ & (\delta(m-n_0)\delta(m+\tau-(N+\nu)-n_1)-\delta(m-(N+\nu)-n_0)\delta(m+\tau-n_1)),&\nonumber\end{eqnarray}
\noindent where $\otimes$ is the convolution operator. Finally, by using the properties of the Kronecker delta function, (\ref{Cor_tr2}) can be re-written as
\begin{eqnarray}
&c_s^{\textrm{(AL)}}(m,\tau)=\sum\limits_{k=-\infty}^{+\infty} \delta\left(m-2k(N+\nu)\right) \otimes \sum\limits_{n_0,n_1=-\nu}^{N+N_W-1}\sigma_{s}^{2}W_{n_0}W_{n_1}\delta\left(mod\left(n_{0}+n_{1},N\right)\right)& \label{Cor_tr3} \\&  (\delta(m-n_0)\delta(\tau-(N+\nu)+n_0-n_1)-\delta(m-(N+\nu)-n_0)\delta(\tau+(N+\nu)+n_0-n_1)).&\nonumber\end{eqnarray}
Finally, from (\ref{Cor_rec1}) and (\ref{Cor_tr3}), (\ref{Cor_AL}) can be easily obtained.
\section*{Appendix B}
\renewcommand{\thesection}{\arabic{section}}
Here, we derive the expression of the CCF for the AL-OFDM signals, which is given in (\ref{CCF_flat2}). Due to the three Kronecker delta functions on the right-hand side of (\ref{CCF_flat}), the two summations over $n_0$ and $n_1$ are taken over a few non-zero terms only. For non-zero terms in (\ref{CCF_flat}), $\delta\left(mod\left(n_{0}+n_{1},N\right)\right)\ne 0$ and either $\delta(\tau-(N+\nu)+n_0-n_1)\ne0$ or $\delta(\tau+(N+\nu)+n_0-n_1)\ne0$. 
\par $\delta\left(mod\left(n_{0}+n_{1},N\right)\right) \ne0$ if and only if $n_0+n_1=qN$, where $q \in  \mathbb{\mathcal{\mathcal{I}}}$. Since $-\nu \le n_0, n_1 \le N+N_W-1$, $\nu < \frac{N}{2}$, and $2(N_W-1)<N$, one can easily see that $-N < -2\nu \le n_0+n_1 \le 2(N+N_W-1)< 3N$, and consequently, $q=0,1,2$. 
As such, we have the following set of constraint linear equations to solve
\begin{equation}
\left\{\begin{array}{l} n_0+n_1=qN,\\n_1-n_0=\tau- \rho (N+\nu),\\
\mbox{subject to} \quad -\nu \le n_0,n_1 \le N+N_W-1, \, q =0,1,2, \,\rho =-1,+1,
\end{array}\right.\label{equations0}
\end{equation}
\noindent where $\rho=+1$ and $-1$ correspond to $\delta(\tau-(N+\nu)+n_0-n_1)$ and $\delta(\tau+(N+\nu)+n_0-n_1)$, respectively.
For given values of $q$ and $\tau$, $n_0$ and $n_1$ are simply obtained from (\ref{equations0}) as
\begin{equation}
n_0=\frac{qN-\tau+\rho(N+\nu)}{2},\label{eq:n0}
\end{equation}
\begin{equation}
n_1=\frac{qN+\tau-\rho(N+\nu)}{2}.\label{eq:n1}
\end{equation}
\par For each value of $q$, the values of $\tau$ are obtained as follows.
\par \textit{A. Values of $\tau$ for $q=0$:} For $q=0$, from the first two equations of (\ref{equations0}), one can easily see that
\begin{equation}
\tau=-2n_0+ \rho (N+\nu).\label{eq:q0}
\end{equation}
 
Since $n_0+n_1=0$ and considering that $-\nu \le n_0,n_1 \le N+N_W-1$, it is straightforward to find that $n_0$ takes integer values in the range $-\nu,...,\nu$. Consequently, based on (\ref{eq:q0}), when $\rho=1$, $\tau \in \{N-\nu,N-\nu+2,...,N+3\nu-2,N+3\nu\}$ and when $\rho=-1$, $\tau \in \{-N-3\nu,-N-3\nu+2,...,-N+\nu-2,-N+\nu\}$. We can compactly write these results as $|\tau| \in \mathbb{\mathcal{\mathcal{I}}}_0$, where $\mathbb{\mathcal{\mathcal{I}}}_0=\{N-\nu,N-\nu+2,...,N+3\nu-2,N+3\nu\}$.
\par \textit{B. Values of $\tau$ for $q=1$:} For $q=1$, from the first two equations of (\ref{equations0}), one can easily see that
\begin{equation}
\tau=N-2n_0+\rho(N+\nu).\label{eq:q1}
\end{equation}
Since $n_0+n_1=N$ and considering that $-\nu \le n_0,n_1 \le N+N_W-1$, it is straightforward to find that $n_0$ takes integer values in the range $1-N_W,...,N+N_W-1$. Consequently, based on (\ref{eq:q1}), when $\rho=1$, $\tau \in \{\nu-2N_W+2,\nu-2N_W+4,...,2N+\nu+2N_W-4,2N+\nu+2N_W-2\}$ and when $\rho=-1$, $\tau \in \{-2N-\nu-2N_W+2,-2N-\nu-2N_W+4,...,-\nu+2N_W-4,-\nu+2N_W-2\}$. We can compactly write these results as $|\tau| \in \mathbb{\mathcal{\mathcal{I}}}_1$, where $\mathbb{\mathcal{\mathcal{I}}}_1=\{\nu-2N_W+2,\nu-2N_W+4,...,2N+\nu+2N_W-4,2N+\nu+2N_W-2\}$.
\par \textit{C. Values of $\tau$ for  $q=2$:} For $q=2$, from the first two equations of (\ref{equations0}), one can easily see that
\begin{equation}
\tau=2N-2n_0+\rho(N+\nu).\label{eq:q2}
\end{equation}
Since $n_0+n_1=2N$ and considering that $-\nu \le n_0,n_1 \le N+N_W-1$, it is straightforward to find that $n_0$ takes integer values in the range $N-N_W+1,...,N+N_W-1$. Consequently, based on (\ref{eq:q2}), when $\rho=1$, $\tau \in \{N+\nu-2N_W+2,N+\nu-2N_W+4,...,N+\nu+2N_W-4,N+\nu+2N_W-2\}$ and when $\rho=-1$, $\tau \in \{-N-\nu-2N_W+2,-N-\nu-2N_W+4,...,-N-\nu+2N_W-4,-N-\nu+2N_W-2\}$. We can compactly write these results as $|\tau| \in \mathbb{\mathcal{\mathcal{I}}}_2$, where $\mathbb{\mathcal{\mathcal{I}}}_2=\{N+\nu-2N_W+2,N+\nu-2N_W+4,...,N+\nu+2N_W-4,N+\nu+2N_W-2\}$.
\par
In a practical STBC-OFDM system, $\nu-2N_W+2 \le N-\nu \le N+\nu-2N_W+2$, $N+\nu+2N_W-2\le N+3\nu \le 2N+\nu+2N_W-2$, and $N$ and $\nu$ are even integers (see, e.g., \cite{WIMAX,book_lte}). Consequently, it is easy to notice that $\mathbb{\mathcal{\mathcal{I}}}_2 \subset \mathbb{\mathcal{\mathcal{I}}}_0 \subset \mathbb{\mathcal{\mathcal{I}}}_1$. As such, when $|\tau| \in \mathbb{\mathcal{\mathcal{I}}}_2$, $q$ takes the previously mentioned three values and hence, there are three non-zero terms in (\ref{CCF_flat}). Furthermore, when $|\tau| \in \mathbb{\mathcal{\mathcal{I}}}_0 \cap \mathbb{\mathcal{\mathcal{I}}}_2^c$, with $\cap$ as the intersection operator and the superscript $c$ as the set complement, $q=0$ or $q=1$, which corresponds to two non-zero terms in (\ref{CCF_flat}). Finally, $q=1$ when $|\tau| \in \mathbb{\mathcal{\mathcal{I}}}_1 \cap \mathbb{\mathcal{\mathcal{I}}}_0^c$, which corresponds to a single non-zero term in (\ref{CCF_flat}). One can also see that when $\rho=1$, $\tau$ is always positive and when $\rho=-1$, $\tau$ is always negative. Therefore, $\rho$ in (\ref{eq:n0}) and (\ref{eq:n1}) can be substituted with $sgn(\tau)$, where $sgn(.)$ is the signum function.
Consequently, based on the above and (\ref{CCF_flat}), one can easily obtain the CCF expression in (\ref{CCF_flat2}).
\section*{Appendix C}
\renewcommand{\thesection}{\arabic{section}}
Here, we analyze the effect of phase noise on the CCF magnitude estimate for AL-OFDM signals. Let
\begin{equation}
 r_{\textrm{PN}}^{(v)}(m)=r^{(v)}(m)e^{-j\phi(m)}, v=0,1, \label{rec_phase}
\end{equation}
\noindent be the $m$th sample of the signal received at the $v$th antenna in the presence of phase noise, where $\phi(m)$ is the phase noise component observed at the $m$th sampling time. This is modeled as a Wiener process, such that $\phi(0)=0$ and $\varphi(m,m_0)=\phi(m)-\phi(m-m_0)$ has a zero-mean Gaussian distribution with variance $\frac{2\pi \beta T m_0}{N+\nu}$, where $\beta$ is the two-sided 3 dB bandwidth of the Lorenzian distribution spectrum and $T$ is the duration of an OFDM symbol \cite{Phasenoise1}.
Using (\ref{est_CCF}), the CCF magnitude is estimated as
\begin{equation}
\hat{C}_{\textrm{PN}}^{\textrm{(AL)}}(\alpha,\tau)=\frac{1}{N_s(N+\nu)}\sum\limits_{m=0}^{N_s(N+\nu)-1}r_{\textrm{PN}}^{(0)}(m)r_{\textrm{PN}}^{(1)}(m+\tau)e^{-j2\pi \alpha m}. \label{est:CCFpn}
\end{equation}
\noindent By following \cite{Test_cyc1}, (\ref{est:CCFpn}) can be further expressed as
\begin{equation}
\hat{C}_{\textrm{PN}}^{\textrm{(AL)}}(\alpha,\tau)=\frac{1}{N_s(N+\nu)}\sum\limits_{m=0}^{N_s(N+\nu)-1}\mathrm{E}\left[r_{\textrm{PN}}^{(0)}(m)r_{\textrm{PN}}^{(1)}(m+\tau)\right]e^{-j2\pi\alpha m} + \varpi(\alpha,\tau), \label{est:CCFpn0}
\end{equation}
where $\varpi(\alpha,\tau)$ is the estimation error which vanishes asymptotically as $N_s\rightarrow\infty$ \cite{Test_cyc1}. Therefore, one can approximate the CCF estimate with its statistical average as
\begin{equation}
\hat{C}_{\textrm{PN}}^{\textrm{(AL)}}(\alpha,\tau) \cong \frac{1}{N_s(N+\nu)}\sum\limits_{m=0}^{N_s(N+\nu)-1}\mathrm{E}\left[r_{\textrm{PN}}^{(0)}(m)r_{\textrm{PN}}^{(1)}(m+\tau)\right]e^{-j2\pi\alpha m}. \label{est:CCFpn1}
\end{equation}
\noindent Without loss of generality, we assume that the received signal starts with the first symbol of an AL block and ends with the second symbol of such a block, and hence, $N_s$ is an even integer. Consequently, (\ref{est:CCFpn1}) can be re-written as
\begin{equation}
\hat{C}_{\textrm{PN}}^{\textrm{(AL)}}(\alpha,\tau) \cong \frac{1}{N_s(N+\nu)}\sum\limits_{k=0}^{\frac{N_s}{2}-1}\sum\limits_{m_1=0}^{2(N+\nu)-1}\mathrm{E}\left[r_{\textrm{PN}}^{(0)}(2k(N+\nu)+m_1)r_{\textrm{PN}}^{(1)}(2k(N+\nu)+m_1+\tau)\right]e^{-j2\pi\alpha m_1}. \label{est:CCFpn2}
\end{equation}
Based on (\ref{rec_phase}), it is easy to express (\ref{est:CCFpn2}) as
\begin{eqnarray}
&\hat{C}_{\textrm{PN}}^{\textrm{(AL)}}(\alpha,\tau) \cong \frac{1}{N_s(N+\nu)}\sum\limits_{k=0}^{\frac{N_s}{2}-1}\sum\limits_{m_1=0}^{2(N+\nu)-1}\mathrm{E}\left[r^{(0)}(2k(N+\nu)+m_1)r^{(1)}(2k(N+\nu)+m_1+\tau)\right]&\nonumber\\ &\times \mathrm{E}\left[e^{-j(\phi(2k(N+\nu)+m_1)+\phi(2k(N+\nu)+m_1+\tau))}\right]e^{-j2\pi\alpha m_1}.\quad\quad\quad\quad\quad& \label{est:CCFpn3}
\end{eqnarray}
Since in the absence of the phase noise, the time-varying cross-correlation function of the received AL-OFDM signal is periodic with the fundamental period $2(N+\nu)$, one can write that
\begin{equation}
c^{\textrm{(AL)}}(m_1,\tau)=\mathrm{E}\left[r^{(0)}(2k(N+\nu)+m_1)r^{(1)}(2k(N+\nu)+m_1+\tau)\right]. \label{cor_npn}
\end{equation}
We assume that the phase noise is approximately constant over each AL block. This assumption can be easily justified as the maximum variance of the phase noise fluctuations over an AL block, which equals $4\pi \beta T$, satisfies $4\pi \beta T \ll 1$, for practical values of $\beta$ and $T$. Consequently, (\ref{est:CCFpn3}) can be re-written as
\begin{equation}
\hat{C}_{\textrm{PN}}^{\textrm{(AL)}}(\alpha,\tau) \cong \frac{1}{N_s(N+\nu)}\sum\limits_{k=0}^{\frac{N_s}{2}-1}\sum\limits_{m_1=0}^{2(N+\nu)-1}c^{\textrm{(AL)}}(m_1,\tau)\mathrm{E}\left[e^{-j2\phi(2k(N+\nu))}\right]e^{-j2\pi\alpha m_1}. \label{est:CCFpn4}
\end{equation}
With $\phi(2k(N+\nu)$ following a Gaussian distribution with mean zero and variance $4\pi \beta kT$, by using the property that $\mathrm{E}\left[e^{-j2\theta}\right]=e^{-2\sigma^2}$, where $\theta \in \mathcal{N}(0,\sigma^2)$\cite{book_prob}, one can easily show that
\begin{equation}
 \hat{C}_{\textrm{PN}}^{\textrm{(AL)}}(\alpha,\tau) \cong \frac{2(1-e^{-4N_s\pi \beta T})}{N_s(1-e^{-8\pi \beta T})}C_{\textrm{PN}}^{\textrm{(AL)}}(\alpha,\tau), \label{est:CCFpn6}
\end{equation}
\noindent i.e., the phase noise scales the CCF magnitude estimate with a factor of $\frac{4(1-e^{-2N_s\pi \beta T})}{N_s(1-e^{-8\pi \beta T})}$.

\newpage

\bibliographystyle{IEEEtran}




\end{document}